\documentclass[10pt,letterpaper,twocolumn,english]{revtex4}
\usepackage{times}
\usepackage[T1]{fontenc}
\usepackage[latin1]{inputenc}
\usepackage{float}
\usepackage{amsmath}
\usepackage[dvips]{graphicx}
\usepackage{amssymb}

\makeatletter
\usepackage{babel}
\makeatother
\begin{document}
\ifx\href\undefined\else\hypersetup{linktocpage=true}\fi 

\title{Is repulsive Casimir force physical?%
\footnote{This paper is based on my Ph.D. dissertation in physics at Virginia
Polytechnic Institute and State University, April (2004) \cite{key-Sung-Nae-Cho}.%
}}

\author{Sung Nae Cho}

\email{sucho2@vt.edu}

\address{Department of Physics, Virginia Polytechnic Institute and State University,
Blacksburg, Virginia  24061, USA}

\date{Monday, August 30, 2004 }

\begin{abstract}
\noindent The Casimir force for charge-neutral, perfect conductors
of non-planar geometric configurations have been investigated. The
configurations are: (1) the plate-hemisphere, (2) the hemisphere-hemisphere
and (3) the spherical shell. The resulting Casimir forces for these
physical arrangements have been found to be attractive. The repulsive
Casimir force found by Boyer for a spherical shell is a special case
requiring stringent material property of the sphere, as well as the
specific boundary conditions for the wave modes inside and outside
of the sphere. The necessary criteria in detecting Boyer's repulsive
Casimir force for a sphere are discussed at the end of this investigation. 
\end{abstract}


\maketitle
\tableofcontents{}

\section{Introduction}

When two electrically neutral, conducting plates are placed parallel
to each other, our understanding from classical electrodynamics tells
us that nothing should happen to these plates. The plates are assumed
to be that made of perfect conductors for simplicity. In 1948, H.
B. G. Casimir and D. Polder faced a similar problem in studying forces
between polarizable neutral molecules in colloidal solutions. Colloidal
solutions are viscous materials, such as paint, that contain micron-sized
particles in a liquid matrix. It had been thought that forces between
such polarizable, neutral molecules were governed by the van der Waals
interaction. The van der Waals interaction is also referred to as
the {}``Lennard-Jones interaction.'' It is a long range electrostatic
interaction that acts to attract two nearby polarizable molecules.
Casimir and Polder found to their surprise that there existed an attractive
force which could not be ascribed to the van der Waals theory. Their
experimental result could not be correctly explained unless the retardation
effect was included in van der Waals' theory. This retarded van der
Waals interaction or Lienard-Wiechert dipole-dipole interaction is
now known as the Casimir-Polder interaction \cite{key-Casimir-Polder,key-Lifshitz,key-Schwinger-DeRaas-Milton,key-Milonni,key-Milton}.
Casimir, following this first work, elaborated on the Casimir-Polder
interaction in predicting the existence of an attractive force between
two electrically neutral, parallel plates of perfect conductors separated
by a small gap \cite{key-Casimir}. This alternative derivation of
the Casimir force is in terms of the difference between the zero-point
energy in vacuum and the zero-point energy in the presence of boundaries.
This force has been confirmed by experiments \cite{key-Lamoreaux,key-Mohideen,key-Bressi-Carugno-Onofrio-Ruoso,key-Yale-Group-Casimir-Polder}
and the phenomenon is what is now known as the {}``Casimir Effect.''
The force responsible for the attraction of two uncharged conducting
plates is accordingly termed the {}``Casimir Force.'' It was shown
later that the Casimir force could be both attractive or repulsive
depending on the geometry and the material property of the conductors
\cite{key-Boyer,key-Maclay,key-Kenneth-Klich-Mann-Revzen}. 

\begin{figure}[H]
\begin{center}\includegraphics[%
  scale=0.5]{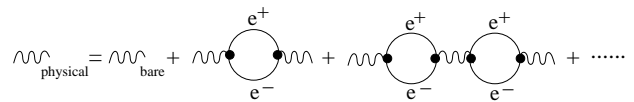}\end{center}

\caption{The vacuum polarization of a photon. \label{cap:vacuum-polarization}}
\end{figure}
 The Casimir effect is regarded as macroscopic manifestation of the
retarded van der Waals interaction between uncharged polarizable molecules
(or atoms). Microscopically, the Casimir effect is due to interactions
between induced multipole moments, where the dipole term is the most
dominant contributor if it is non-vanishing. Therefore, the dipole
interaction is exclusively referred to, unless otherwise explicitly
stated, throughout this investigation. The induced dipole moments
can be qualitatively explained by the concept of {}``vacuum polarization''
in quantum electrodynamics (QED). The idea is that a photon, whether
real or virtual, has a charged particle content. Namely, the internal
loop, illustrated in Figure \ref{cap:vacuum-polarization}, can be
$e^{+}e^{-},$ $\mu^{+}\mu^{-},$ $\tau^{+}\tau^{-},$ $\pi^{+}\pi^{-}$
or $q\bar{q}$ pairs, etc. Its correctness have been born out from
the precision measurements of Lamb shift \cite{key-Lundeen-Pipkin,key-Dubler-et-al}
and photoproduction \cite{key-photo-pro-M-Ya-Amusia,key-photo-pro-F-J-Gilman,key-photo-pro-Lautrup-Peterman-Rafael,key-photo-pro-Krawczyk-Zembrzuski-Staszel}
experiments over a vast range of energies. For the almost zero energy
photons considered in the Casimir effect, these pairs last for a time
interval $\triangle t$ consistent with that given by the Heisenberg
uncertainty principle $\triangle E\cdot\triangle t=h,$ where $\triangle E$
is the energy imbalance and $h$ is the Planck constant. These virtual
charged particles can induce the requisite polarizability on the boundary
of the dielectric (or conducting) plates which explained the Casimir
effect. However, the dipole strength is left as a free parameter in
the calculations because it cannot be readily calculated \cite{key-Uehling}.
Its value can be determined from experiments. 

Once this idea is taken for granted, one can then move forward to
calculate the effective, temperature averaged, energy due to the dipole-dipole
interactions with the time retardation effect folded in. The energy
between the dielectric (or conducting) media is obtained from the
allowed modes of electromagnetic waves determined by the Maxwell equations
together with the boundary conditions. The Casimir force is then obtained
by taking the negative gradient of the energy in space. This approach,
as opposed to full atomistic treatment of the dielectrics (or conductors),
is justified as long as the most significant photon wavelengths determining
the interaction are large when compared with the spacing of the lattice
points in the media. The effect of all the multiple dipole scattering
by atoms in the dielectric (or conducting) media simply enforces the
macroscopic reflection laws of electromagnetic waves. For instance,
in the case of the two parallel plates, the most significant wavelengths
are those of the order of the plate gap distance. When this wavelength
is large compared with the interatomic distances, the macroscopic
electromagnetic theory can be used with impunity. The geometric configuration
can introduce significant complications, which is the subject matter
this study is going to address.

In order to handle the dipole-dipole interaction Hamiltonian in this
case, the classical electromagnetic fields have to be quantized into
the photon representation first. The photon with non-zero occupation
number have energies in units of $\hbar\omega;$ where $\hbar$ is
the Planck constant divided by $2\pi,$ and $\omega,$ the angular
frequency. The lowest energy state of the electromagnetic fields has
energy $\hbar\omega/2.$ They are called the vacuum or the zero point
energy state, and they play a major role in the Casimir effect. Throughout
this investigation, the terminology {}``photon'' is used to represent
the entity with energy $\hbar\omega,$ or the entity with energy $\hbar\omega/2$
unless explicitly stated otherwise. With this in mind, the quantized
field energy is written as 

\begin{align}
\mathcal{H}'_{n_{s},b} & =\left[n_{s}+\frac{1}{2}\right]\hbar c\Theta_{k'}\nonumber \\
 & \times\sum_{n_{1}=0}^{\infty}\sum_{n_{2}=0}^{\infty}\sum_{n_{3}=0}^{\infty}\sqrt{\sum_{i=1}^{3}\left[k'_{i}\left(n_{i},L_{i}\right)\right]^{2}},\label{eq:stationary-state-energy-bounded}\end{align}
 where $c$ is the speed of light in empty space, $\Theta_{k'}$ is
the degree of freedom in polarization, $k'_{i}$ is the wave number,
$n_{i}$ is the wave mode number and $L_{i},$ the boundary length.
The subscript $b$ of $\mathcal{H}'_{n_{s},b}$ denotes the bounded
space. For electromagnetic waves, the degree of freedom in polarization,
$\Theta_{k'}\equiv2.$ Similarly, in free space, the field energy
is quantized in the form \begin{align}
\mathcal{H}'_{n_{s},u}\equiv\mathcal{H}'_{n_{s}} & =\frac{\left[n_{s}+\frac{1}{2}\right]\hbar c\Theta_{k'}}{f_{1}\left(L_{1}\right)f_{2}\left(L_{2}\right)f_{3}\left(L_{3}\right)}\int_{0}^{\infty}\int_{0}^{\infty}\nonumber \\
 & \times\int_{0}^{\infty}\sqrt{\sum_{i=1}^{3}\left[k'_{i}\left(n_{i},L_{i}\right)\right]^{2}}dk'_{1}dk'_{2}dk'_{3},\label{eq:stationary-state-energy-unbounded}\end{align}
 where the subscript $u$ of $\mathcal{H}'_{n_{s},u}$ denotes free
(unbounded) space and the functional $f_{i}\left(L_{i}\right)$ in
the denominator is equal to $\zeta_{zero}n_{i}^{-1}L_{i}^{-1}$ for
a given $L_{i}.$ Here $\zeta_{zero}$ is the zeroes of the function
representing the transversal component of the electric field. The
corresponding zero point energy for bounded (or unbounded) space is
obtained by setting $n_{s}=0$ in equations (\ref{eq:stationary-state-energy-bounded})
and (\ref{eq:stationary-state-energy-unbounded}), respectively.

\section{Reflection Dynamics}

In principle, the atomistic approach utilizing the Casimir-Polder
interaction \cite{key-Casimir-Polder} explains the Casimir effect
observed in any system. Unfortunately, the pairwise summation of the
intermolecular forces for systems containing large number of atoms
can become very complicated. H. B. G. Casimir, realizing the linear
relationship between the field and the polarization, devised an easier
approach to the calculation of Casimir effect for large systems such
as two perfectly conducting parallel plates \cite{key-Casimir}. The
Casimir effect have been also explained by Schwinger utilizing his
original invention, {}``Source Theory'' \cite{key-Milton,key-Schwinger-DeRaas-Milton}.

In this investigation, we do not follow Casimir's energy method, nor
do we take the route of Schwinger's source theory. Instead, we adopt
the vacuum pressure approach introduced by Milonni, Cook and Goggin
\cite{key-Milonni-Cook-Goggin}, which is a simple elaboration on
Casimir's original calculation technique utilizing the boundary conditions.
We choose to consider the vacuum pressure approach over both Casimir's
energy method and Schwinger's source theory not because it is a superior
technique, but simply because it is the easiest one for the physical
arrangements considered in this investigation. 

The three physical arrangements for the boundary configurations considered
in this investigation are: (1) the plate-hemisphere, (2) the hemisphere-hemisphere
and (3) a sphere formed by brining two hemispheres together. Because
the geometric configurations of items (2) and (3) are special versions
of the more general, plate-hemisphere configuration, the basic reflection
dynamics needed for the plate-hemisphere case is worked out first.
The results can then be applied to the hemisphere-hemisphere and the
sphere configurations later. 

\begin{figure}[t]
\begin{center}\includegraphics[%
  scale=0.4]{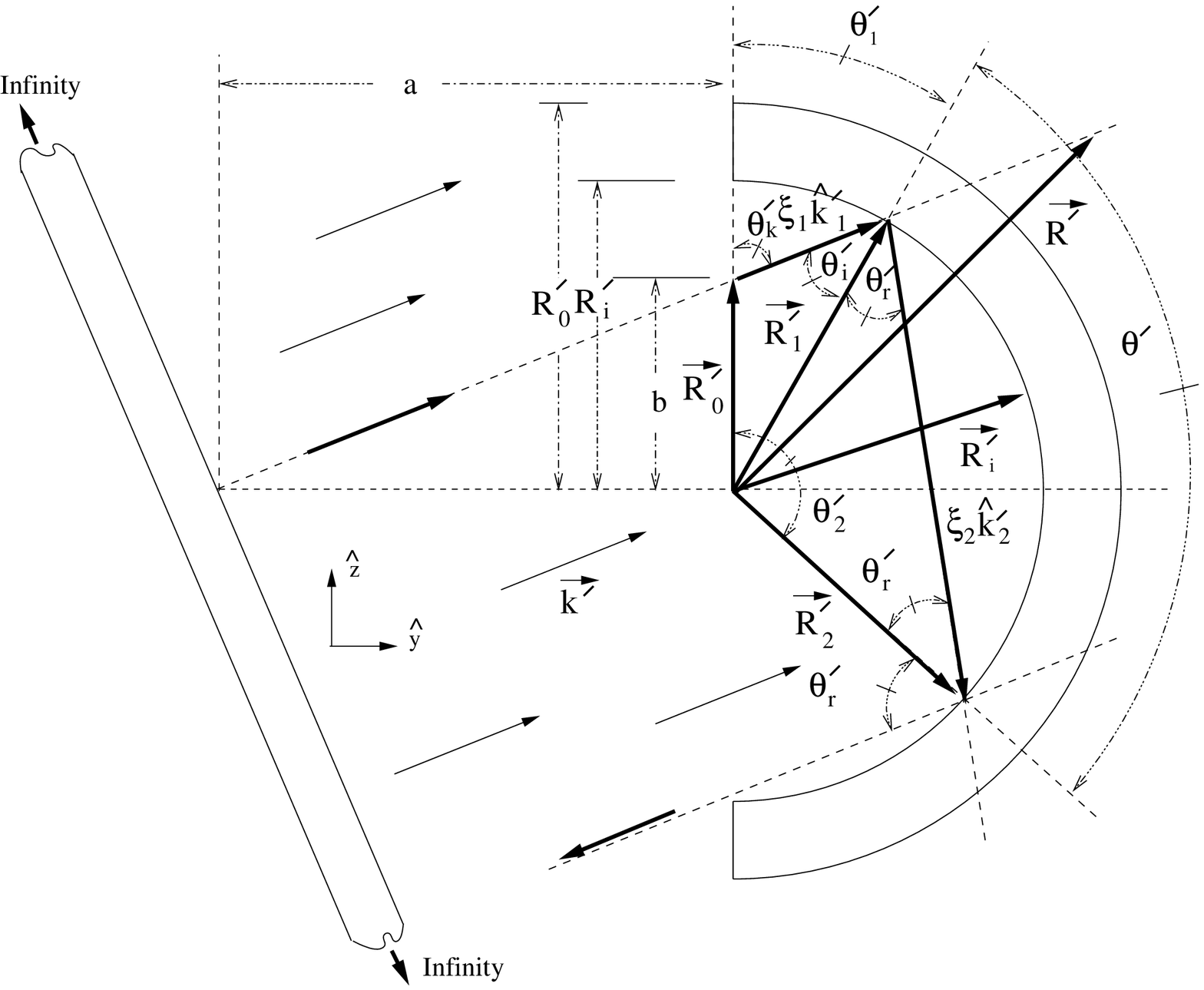}\end{center}

\caption{The plane of incidence view of plate-hemisphere configuration. The
waves that are allowed through internal reflections in the hemisphere
cavity must satisfy the relation $\lambda\leq2\left\Vert \vec{R'}_{2}-\vec{R'}_{1}\right\Vert .$
\label{cap:cross-sectional-view-plate-and-hemisphere}}
\end{figure}

The vacuum-fields are subject to the appropriate boundary conditions.
For boundaries made of perfect conductors, the transverse components
of the electric field are zero at the surface. For this simplification,
the skin depth of penetration is considered to be zero. The plate-hemisphere
under consideration is shown in Figure \ref{cap:cross-sectional-view-plate-and-hemisphere}.
The solutions to the vacuum-fields are that of the Cartesian version
of the free Maxwell field vector potential differential equation $\nabla^{2}\vec{A}\left(\vec{R}\right)-c^{-2}\partial_{t}^{2}\vec{A}\left(\vec{R}\right)=0,$
where the Coulomb gauge $\vec{\nabla}\cdot\vec{A}=0$ and the absence
of the source $\rho$ for scalar potential $\Phi$ have been imposed,
$\Phi\left(\rho,\left\Vert \vec{R}\right\Vert \right)=0.$ The electric
and the magnetic field component of the vacuum-field are given by
$\vec{E}=-c^{-1}\partial_{t}\vec{A}$ and $\vec{B}=\vec{\nabla}\times\vec{A},$
where $\vec{A}$ is the free field vector potential. The vanishing
of the transversal component of the electric field at the perfect
conductor surface implies that the solution for $\vec{E}$ is in the
form of $\vec{E}\propto\sin\left(2\pi\lambda^{-1}\left\Vert \vec{L}\right\Vert \right),$
where $\lambda$ is the wavelength and $\left\Vert \vec{L}\right\Vert $
is the path length between the boundaries. The wavelength is restricted
by the condition $\lambda\leq2\left\Vert \vec{R'}_{2}-\vec{R'}_{1}\right\Vert \equiv2\xi_{2},$
where $\vec{R'}_{2}$ and $\vec{R'}_{1}$ are two immediate reflection
points in the hemisphere cavity of Figure \ref{cap:cross-sectional-view-plate-and-hemisphere}.
In order to compute the modes allowed inside the hemisphere resonator,
a detailed knowledge of the reflections occurring in the hemisphere
cavity is needed. This is described in the following section.

\subsection{Reflection Points on the Surface of a Resonator}

The wave vector directed along an arbitrary direction in Cartesian
coordinates is written as \begin{align}
\vec{k'}_{1}\left(k'_{1,x},k'_{1,y},k'_{1,z}\right) & =\sum_{i=1}^{3}k'_{1,i}\hat{e_{i}},\label{eq:arbitrary-k-vector-NA}\end{align}
 where \begin{align*}
k'_{1,i} & =\left\{ \begin{array}{cc}
i=1\rightarrow k'_{1,x}, & \hat{e_{1}}=\hat{x},\\
\\i=2\rightarrow k'_{1,y}, & \hat{e_{2}}=\hat{y},\\
\\i=3\rightarrow k'_{1,z}, & \hat{e_{3}}=\hat{z}.\end{array}\right.\end{align*}
 Hence, the unit wave vector, $\hat{k'}_{1}=\left\Vert \vec{k'}_{1}\right\Vert ^{-1}\sum_{i=1}^{3}k'_{1,i}\hat{e_{i}}.$
Define the initial position $\vec{R'}_{0}$ for the incident wave
$\vec{k'}_{1},$ \begin{align}
\vec{R'}_{0}\left(r'_{0,x},r'_{0,y},r'_{0,z}\right) & =\sum_{i=1}^{3}r'_{0,i}\hat{e_{i}},\label{eq:initial-arbitrary-k-vector-position-NA}\end{align}
 where \[
r'_{0,i}=\left\{ \begin{array}{c}
i=1\rightarrow r'_{0,x},\\
\\i=2\rightarrow r'_{0,y},\\
\\i=3\rightarrow r'_{0,z}.\end{array}\right.\]
 Here it should be noted that $\vec{R'}_{0}$ really has only components
$r'_{0,x}$ and $r'_{0,z}.$ But nevertheless, one can always set
$r'_{0,y}=0$ whenever needed. Since no particular wave vectors with
specified wave lengths are prescribed initially, it is desirable to
employ a parameterization scheme to represent these wave vectors.
The line segment traced out by this wave vector $\hat{k'}_{1}$ is
formulated in the parametric form \begin{align}
\vec{R'}_{1} & =\xi_{1}\hat{k'}_{1}+\vec{R'}_{0}\nonumber \\
 & =\sum_{i=1}^{3}\left[r'_{0,i}+\xi_{1}\left\Vert \vec{k'}_{1}\right\Vert ^{-1}k'_{1,i}\right]\hat{e_{i}},\label{eq:parametrized-line-segment-generalized-NA}\end{align}
 where the variable $\xi_{1}$ is a positive definite parameter. Here
$\vec{R'}_{1}$ is the first reflection point on the hemisphere. In
terms of spherical coordinate variables, $\vec{R'}_{1}$ takes the
form \begin{align}
\vec{R'}_{1}\left(r'_{i},\theta'_{1},\phi'_{1}\right) & =r'_{i}\sum_{i=1}^{3}\Lambda'_{1,i}\hat{e_{i}},\label{eq:sphere-vector-generalized-NA}\end{align}
 where \[
\left\{ \begin{array}{c}
\Lambda'_{1,1}=\sin\theta'_{1}\cos\phi'_{1},\\
\\\Lambda'_{1,2}=\sin\theta'_{1}\sin\phi'_{1},\\
\\\Lambda'_{1,3}=\cos\theta'_{1}.\qquad\;\:\end{array}\right.\]
 Here $r'_{i}$ is the hemisphere radius, $\theta'_{1}$ and $\phi'_{1}$
are the polar and the azimuthal angle respectively of $\vec{R'}_{1}$
at the first reflection point. Notice that subscript $i$ of $r'_{i}$
denotes {}``inner radius'' not a summation index. 

By combining equations (\ref{eq:parametrized-line-segment-generalized-NA})
and (\ref{eq:sphere-vector-generalized-NA}), we can solve for the
parameter $\xi_{1}.$ It can be shown that \begin{align}
\xi_{1} & \equiv\xi_{1,p}\nonumber \\
 & =-\hat{k'}_{1}\cdot\vec{R'}_{0}+\sqrt{\left[\hat{k'}_{1}\cdot\vec{R'}_{0}\right]^{2}+\left[r'_{i}\right]^{2}-\left\Vert \vec{R'}_{0}\right\Vert ^{2}},\label{eq:positive-root-k-NA}\end{align}
 where the positive root for $\xi_{1}$ have been chosen due to the
restriction $\xi_{1}\geq0.$ The detailed proof of equation (\ref{eq:positive-root-k-NA})
is given in Appendix A of reference \cite{key-Sung-Nae-Cho}. Substituting
$\xi_{1}$ in equation (\ref{eq:parametrized-line-segment-generalized-NA}),
the first reflection point off the inner hemisphere surface is expressed
as \begin{align}
\vec{R'}_{1}\left(\xi_{1,p};\vec{R'}_{0},\hat{k'}_{1}\right) & =\sum_{i=1}^{3}\left[r'_{0,i}+\xi_{1,p}\left\Vert \vec{k'}_{1}\right\Vert ^{-1}k'_{1,i}\right]\hat{e_{i}},\label{eq:1st-bounce-off-point-NA}\end{align}
 where $\xi_{1,p}$ is from equation (\ref{eq:positive-root-k-NA}). 

The incoming wave vector $\vec{k'}_{i}$ can always be decomposed
into parallel and perpendicular components, $\vec{k'}_{i,\parallel}$
and $\vec{k'}_{i,\perp},$ with respect to the local reflection surface.
It is shown in Appendix A of reference \cite{key-Sung-Nae-Cho} that
the reflected wave vector $\vec{k'}_{r}$ has the form \begin{align*}
\vec{k'}_{r} & =\alpha_{r,\perp}\left[\hat{n'}\times\vec{k'}_{i}\right]\times\hat{n'}-\alpha_{r,\parallel}\hat{n'}\cdot\vec{k'}_{i}\hat{n'},\end{align*}
 where the quantities $\alpha_{r,\parallel}$ and $\alpha_{r,\perp}$
are the reflection coefficients and $\hat{n'}$ is a unit surface
normal. For the perfect reflecting surfaces, $\alpha_{r,\parallel}=\alpha_{r,\perp}=1.$
In component form, \begin{align*}
\vec{k'}_{r} & =\sum_{l=1}^{3}\left\{ \alpha_{r,\perp}\left[n'_{n}k'_{i,l}n'_{n}-n'_{l}k'_{i,n}n'_{n}\right]\right.\\
 & \left.-\alpha_{r,\parallel}n'_{n}k'_{i,n}n'_{l}\right\} \hat{e_{l}},\end{align*}
 where it is understood that $\hat{n'}$ is already normalized and
Einstein summation convention is applied to the index $n.$ The second
reflection point $\vec{R'}_{2}$ is found then by repeating the steps
done for $\vec{R'}_{1}$ and by using the expression $\vec{k'}_{r}\equiv\vec{k'}_{r}/\left\Vert \vec{k'}_{r}\right\Vert ,$
\begin{align*}
\vec{R'}_{2} & =\vec{R'}_{1}+\xi_{2,p}\hat{k'}_{r}\\
 & =\vec{R'}_{1}+\xi_{2,p}\frac{\alpha_{r,\perp}\left[\hat{n'}\times\vec{k'}_{i}\right]\times\hat{n'}-\alpha_{r,\parallel}\hat{n'}\cdot\vec{k'}_{i}\hat{n'}}{\left\Vert \alpha_{r,\perp}\left[\hat{n'}\times\vec{k'}_{i}\right]\times\hat{n'}-\alpha_{r,\parallel}\hat{n'}\cdot\vec{k'}_{i}\hat{n'}\right\Vert },\end{align*}
 where $\xi_{2,p}$ is the new positive definite parameter for the
second reflection point. 

The incidence plane of reflection is determined solely by the incident
wave $\vec{k'}_{i}$ and the local normal $\vec{n'}_{i}$ of the reflecting
surface. It is important to recognize the fact that the subsequent
successive reflections of this incoming wave will be confined to this
particular incidence plane. This incident plane can be characterized
by a unit normal vector. For the system shown in Figure \ref{cap:cross-sectional-view-plate-and-hemisphere},
\begin{align*}
\vec{k'}_{i} & =\vec{k'}_{1},\end{align*}
 and \begin{align*}
\vec{n'}_{n'_{i},1} & =-\xi_{1,p}\hat{k'}_{1}-\vec{R'}_{0}.\end{align*}
The unit vector which represents the incidence plane is given by \begin{align*}
\hat{n'}_{p,1} & =-\left\Vert \vec{n'}_{p,1}\right\Vert ^{-1}\sum_{i=1}^{3}\epsilon_{ijk}k'_{1,j}r'_{0,k}\hat{e_{i}},\end{align*}
 where the summations over indices $j$ and $k$ are implicit. If
the plane of incidence is represented by a scalar function $f\left(x',y',z'\right),$
then its unit normal vector $\hat{n'}_{p,1}$ will satisfy the relationship
$\hat{n'}_{p,1}\propto\vec{\nabla'}f_{p,1}\left(x',y',z'\right).$
It can be shown (see Appendix A of reference \cite{key-Sung-Nae-Cho})
that \begin{align}
f_{p,1}\left(\nu'_{1},\nu'_{2},\nu'_{3}\right) & =-\left\Vert \vec{n'}_{p,1}\right\Vert ^{-1}\sum_{i=1}^{3}\epsilon_{ijk}k'_{1,j}r'_{0,k}\nu'_{i},\label{eq:f-p-i-plane-of-incidence-NA}\end{align}
 where \[
i=\left\{ \begin{array}{c}
1\rightarrow\nu'_{1}=x',\\
\\2\rightarrow\nu'_{2}=y',\\
\\3\rightarrow\nu'_{3}=z'\end{array}\right.\]
 and \[
-\infty\leq\left\{ \nu'_{1}=x',\nu'_{2}=y',\nu'_{3}=z'\right\} \leq\infty.\]

\begin{figure}[t]
\begin{center}\includegraphics[%
  scale=0.5]{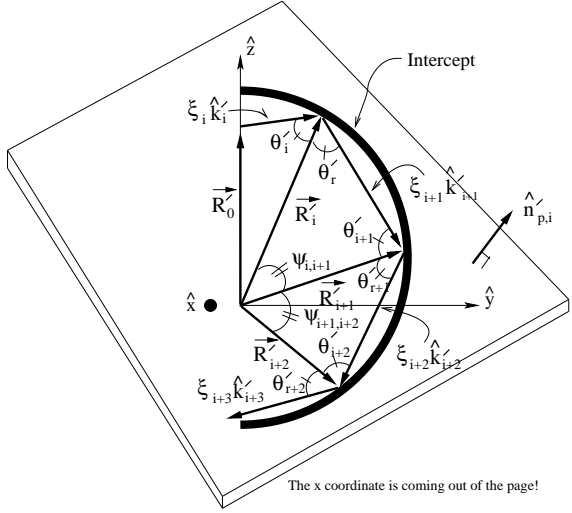}\end{center}

\caption{The thick line shown here represents the intersection between hemisphere
surface and the plane of incidence. The unit vector normal to the
plane of incidence is given by $\hat{n'}_{p,1}=-\left\Vert \vec{n'}_{p,1}\right\Vert ^{-1}\sum_{i=1}^{3}\epsilon_{ijk}k'_{1,j}r'_{0,k}\hat{e_{i}}.$
\label{cap:plane-sphere-intersection}}
\end{figure}

The surface of a sphere or hemisphere is defined through the relation
\begin{align*}
f_{hemi}\left(x',y',z'\right) & =\left[r'_{i}\right]^{2}-\sum_{i=1}^{3}\left[\nu'_{i}\right]^{2},\end{align*}
 where $r'_{i}$ is the radius of sphere and the subscript $i$ denotes
the inner surface. The intercept of interest is shown in Figure \ref{cap:plane-sphere-intersection}.
The intersection between the hemisphere surface and the incidence
plane $f_{p,1}\left(\nu'_{1},\nu'_{2},\nu'_{3}\right)$ is given through
the relation \begin{align*}
f_{hemi}\left(x',y',z'\right)-f_{p,1}\left(x',y',z'\right) & =0.\end{align*}
 After substitution of $f_{p,1}\left(x',y',z'\right)$ and $f_{hemi}\left(x',y',z'\right),$
we have \begin{align*}
\sum_{i=1}^{3}\left\{ \left[\nu'_{i}\right]^{2}-\left\Vert \vec{n'}_{p,1}\right\Vert ^{-1}\epsilon_{ijk}k'_{1,j}r'_{0,k}\nu'_{i}\right\} -\left[r'_{i}\right]^{2} & =0,\end{align*}
 where \[
i=\left\{ \begin{array}{c}
1\rightarrow\nu'_{1}=x',\\
\\2\rightarrow\nu'_{2}=y',\\
\\3\rightarrow\nu'_{3}=z'.\end{array}\right.\]
 The term $\left[r'_{i}\right]^{2}$ can be rewritten in the form
\begin{align*}
\left[r'_{i}\right]^{2} & =\sum_{i=1}^{3}\left[r'_{i,i}\right]^{2},\end{align*}
 where \[
\left\{ \begin{array}{c}
r'_{i,1}=r'_{i,x'},\\
\\r'_{i,2}=r'_{i,y'},\\
\\r'_{i,3}=r'_{i,z'}.\end{array}\right.\]
 Solving for $\nu'_{i},$ it can be shown (see Appendix A of reference
\cite{key-Sung-Nae-Cho}) that \begin{align}
\nu'_{i} & =\frac{1}{2}\left\Vert \vec{n'}_{p,1}\right\Vert ^{-1}\epsilon_{ijk}k'_{1,j}r'_{0,k}\nonumber \\
 & \pm\sqrt{\left[\frac{1}{2}\left\Vert \vec{n'}_{p,1}\right\Vert ^{-1}\epsilon_{ijk}k'_{1,j}r'_{0,k}\right]^{2}+\left[r'_{i,i}\right]^{2}},\label{eq:hemi-plane-intercept-root-NA}\end{align}
 where $i=1,2,3$ and $\epsilon_{ijk}$ is the Levi-Civita symbol.
The result for $\nu'_{i}$ shown above provide a set of discrete reflection
points found by the intercept between the hemisphere and the plane
of incidence. Using spherical coordinate representations for the variables
$r'_{i,1},$ $r'_{i,2}$ and $r'_{i,3},$ \[
\left\{ \begin{array}{c}
r'_{i,1}=r'_{i}\sin\theta'\cos\phi',\\
\\r'_{i,2}=r'_{i}\sin\theta'\sin\phi',\\
\\r'_{i,3}=r'_{i}\cos\theta',\qquad\;\:\end{array}\right.\]
 the initial reflection point $\vec{R'}_{1}$ can be expressed in
terms of the spherical coordinate variables $\left(r'_{i},\theta'_{1},\phi'_{1}\right),$
\begin{align}
\vec{R'}_{1}\left(r'_{i},\theta'_{1},\phi'_{1}\right) & =\sum_{i=1}^{3}\nu'_{1,i}\left(r'_{i},\theta'_{1},\phi'_{1}\right)\hat{e_{i}},\label{eq:1st-Reflection-Point-NA}\end{align}
 where \[
i=\left\{ \begin{array}{c}
1\rightarrow\nu'_{1,1}=r'_{i}\sin\theta'_{1}\cos\phi'_{1},\\
\\2\rightarrow\nu'_{1,2}=r'_{i}\sin\theta'_{1}\sin\phi'_{1},\\
\\3\rightarrow\nu'_{1,3}=r'_{i}\cos\theta'_{1}.\qquad\;\:\end{array}\right.\]
 Here $r'_{i}$ is the hemisphere radius, $\phi'_{1}$ and $\theta'_{1},$
the polar and azimuthal angle, respectively. The angles $\phi'_{1}$
and $\theta'_{1}$ are found to be (see Appendix A of reference \cite{key-Sung-Nae-Cho}):
\begin{equation}
\left\{ \begin{array}{c}
\lim_{\varepsilon\rightarrow0}\left(0\leq\phi'_{1}\leq\frac{1}{2}\pi-\left|\varepsilon\right|\right),\\
\\\lim_{\varepsilon\rightarrow0}\left(\pi+\left|\varepsilon\right|\leq\phi'_{1}\leq\frac{3}{2}\pi-\left|\varepsilon\right|\right),\\
\\\phi'_{1}=\arctan\left(\sqrt{\frac{\Upsilon_{1,2}^{2}-\Upsilon_{1,2}\left\Vert \vec{n'}_{p,1}\right\Vert ^{-1}\epsilon_{2mn}k'_{1,m}r'_{0,n}}{\Upsilon_{1,1}^{2}-\Upsilon_{1,1}\left\Vert \vec{n'}_{p,1}\right\Vert ^{-1}\epsilon_{1jk}k'_{1,j}r'_{0,k}}}\right);\end{array}\right.\label{eq:phi1-prime-0-90-and-180-270-final}\end{equation}
\begin{equation}
\left\{ \begin{array}{c}
\lim_{\varepsilon\rightarrow0}\left(\frac{1}{2}\pi+\left|\varepsilon\right|\leq\phi'_{1}\leq\pi-\varepsilon\right),\\
\\\lim_{\varepsilon\rightarrow0}\left(\frac{3}{2}\pi+\left|\varepsilon\right|\leq\phi'_{1}<2\pi-\left|\varepsilon\right|\right),\\
\\\phi'_{1}=\arctan\left(-\sqrt{\frac{\Upsilon_{1,2}^{2}-\Upsilon_{1,2}\left\Vert \vec{n'}_{p,1}\right\Vert ^{-1}\epsilon_{2mn}k'_{1,m}r'_{0,n}}{\Upsilon_{1,1}^{2}-\Upsilon_{1,1}\left\Vert \vec{n'}_{p,1}\right\Vert ^{-1}\epsilon_{1jk}k'_{1,j}r'_{0,k}}}\right);\end{array}\right.\label{eq:phi1-prime-90-180-and-270-360-final}\end{equation}
\begin{equation}
\left\{ \begin{array}{c}
\lim_{\varepsilon\rightarrow0}\left(0\leq\theta'_{1}\leq\frac{1}{2}\pi-\left|\varepsilon\right|\right),\\
\\\theta'_{1}=\arctan\left(\left[\frac{\Upsilon_{1,1}^{2}+\Upsilon_{1,2}^{2}}{\Upsilon_{1,3}^{2}-\Upsilon_{1,3}\left\Vert \vec{n'}_{p,1}\right\Vert ^{-1}\epsilon_{3qr}k'_{1,q}r'_{0,r}}\right.\right.\\
\left.\left.-\frac{\Upsilon_{1,1}\epsilon_{1jk}k'_{1,j}r'_{0,k}+\Upsilon_{1,2}\epsilon_{2mn}k'_{1,m}r'_{0,n}}{\left\Vert \vec{n'}_{p,1}\right\Vert \left\{ \Upsilon_{1,3}^{2}-\Upsilon_{1,3}\left\Vert \vec{n'}_{p,1}\right\Vert ^{-1}\epsilon_{3qr}k'_{1,q}r'_{0,r}\right\} }\right]^{1/2}\right);\end{array}\right.\label{eq:theta1-prime-0-to-90-degree-final}\end{equation}
\begin{equation}
\left\{ \begin{array}{c}
\lim_{\varepsilon\rightarrow0}\left(\frac{1}{2}\pi+\left|\varepsilon\right|\leq\theta'_{1}\leq\pi\right),\\
\\\theta'_{1}=\arctan\left(-\left[\frac{\Upsilon_{1,1}^{2}+\Upsilon_{1,2}^{2}}{\Upsilon_{1,3}^{2}-\Upsilon_{1,3}\left\Vert \vec{n'}_{p,1}\right\Vert ^{-1}\epsilon_{3qr}k'_{1,q}r'_{0,r}}\right.\right.\\
\left.\left.-\frac{\Upsilon_{1,1}\epsilon_{1jk}k'_{1,j}r'_{0,k}+\Upsilon_{1,2}\epsilon_{2mn}k'_{1,m}r'_{0,n}}{\left\Vert \vec{n'}_{p,1}\right\Vert \left\{ \Upsilon_{1,3}^{2}-\Upsilon_{1,3}\left\Vert \vec{n'}_{p,1}\right\Vert ^{-1}\epsilon_{3qr}k'_{1,q}r'_{0,r}\right\} }\right]^{1/2}\right),\end{array}\right.\label{eq:theta1-prime-90-to-180-degree-final}\end{equation}
 where \begin{eqnarray*}
\Upsilon_{1,i}=r'_{0,i}+\xi_{1,p}\left\Vert \vec{k'}_{1}\right\Vert ^{-1}k'_{1,i}, &  & i=1,2,3.\end{eqnarray*}
 Similarly, the second reflection point on the inner hemisphere surface
is given by (see Appendix A of reference \cite{key-Sung-Nae-Cho}):
\begin{align}
\vec{R'}_{2}\left(r'_{i},\theta'_{2},\phi'_{2}\right) & =\sum_{i=1}^{3}\nu'_{2,i}\left(r'_{i},\theta'_{2},\phi'_{2}\right)\hat{e_{i}},\label{eq:2nd-Reflection-Point-NA}\end{align}
 where \[
i=\left\{ \begin{array}{c}
1\rightarrow\nu'_{2,1}=r'_{i}\sin\theta'_{2}\cos\phi'_{2},\\
\\2\rightarrow\nu'_{2,2}=r'_{i}\sin\theta'_{2}\sin\phi'_{2},\\
\\3\rightarrow\nu'_{2,3}=r'_{i}\cos\theta'_{2}.\qquad\;\:\end{array}\right.\]
 Here the spherical angles $\phi'_{2}$ and $\theta'_{2}$ are found
to be (see Appendix A of reference \cite{key-Sung-Nae-Cho}): \begin{equation}
\left\{ \begin{array}{c}
\lim_{\varepsilon\rightarrow0}\left(0\leq\phi'_{2}\leq\frac{1}{2}\pi-\left|\varepsilon\right|\right),\\
\\\lim_{\varepsilon\rightarrow0}\left(\pi+\left|\varepsilon\right|\leq\phi'_{2}\leq\frac{3}{2}\pi-\left|\varepsilon\right|\right),\\
\\\phi'_{2}=\arctan\left(\left\{ \left[\frac{1}{2}\left\Vert \vec{n'}_{p,1}\right\Vert ^{-1}\epsilon_{2mn}k'_{1,m}r'_{0,n}-\grave{\nu}'_{2,2}\right]^{2}\right.\right.\\
\left.-\left[\frac{1}{2}\left\Vert \vec{n'}_{p,1}\right\Vert ^{-1}\epsilon_{2mn}k'_{1,m}r'_{0,n}\right]^{2}\right\} ^{1/2}\\
\times\left\{ \left[\frac{1}{2}\left\Vert \vec{n'}_{p,1}\right\Vert ^{-1}\epsilon_{1jk}k'_{1,j}r'_{0,k}-\grave{\nu}'_{2,1}\right]^{2}\right.\\
\left.\left.-\left[\frac{1}{2}\left\Vert \vec{n'}_{p,1}\right\Vert ^{-1}\epsilon_{1jk}k'_{1,j}r'_{0,k}\right]^{2}\right\} ^{-1/2}\right);\end{array}\right.\label{eq:phi2-prime-0-90-and-180-270-final}\end{equation}
 \begin{equation}
\left\{ \begin{array}{c}
\lim_{\varepsilon\rightarrow0}\left(\frac{1}{2}\pi+\left|\varepsilon\right|\leq\phi'_{2}\leq\pi-\varepsilon\right),\\
\\\lim_{\varepsilon\rightarrow0}\left(\frac{3}{2}\pi+\left|\varepsilon\right|\leq\phi'_{2}<2\pi-\left|\varepsilon\right|\right),\\
\\\phi'_{2}=\arctan\left(-\left\{ \left[\frac{1}{2}\left\Vert \vec{n'}_{p,1}\right\Vert ^{-1}\epsilon_{2mn}k'_{1,m}r'_{0,n}-\grave{\nu}'_{2,2}\right]^{2}\right.\right.\\
\left.-\left[\frac{1}{2}\left\Vert \vec{n'}_{p,1}\right\Vert ^{-1}\epsilon_{2mn}k'_{1,m}r'_{0,n}\right]^{2}\right\} ^{1/2}\\
\times\left\{ \left[\frac{1}{2}\left\Vert \vec{n'}_{p,1}\right\Vert ^{-1}\epsilon_{1jk}k'_{1,j}r'_{0,k}-\grave{\nu}'_{2,1}\right]^{2}\right.\\
\left.\left.-\left[\frac{1}{2}\left\Vert \vec{n'}_{p,1}\right\Vert ^{-1}\epsilon_{1jk}k'_{1,j}r'_{0,k}\right]^{2}\right\} ^{-1/2}\right);\end{array}\right.\label{eq:phi2-prime-90-180-and-270-360-final}\end{equation}
 \begin{equation}
\left\{ \begin{array}{c}
\lim_{\varepsilon\rightarrow0}\left(0\leq\theta'_{2}\leq\frac{1}{2}\pi-\left|\varepsilon\right|\right),\\
\\\theta'_{2}=\arctan\left(\left\{ \left[\frac{1}{2}\left\Vert \vec{n'}_{p,1}\right\Vert ^{-1}\epsilon_{1jk}k'_{1,j}r'_{0,k}-\grave{\nu}'_{2,1}\right]^{2}\right.\right.\\
-\left[\frac{1}{2}\left\Vert \vec{n'}_{p,1}\right\Vert ^{-1}\epsilon_{1jk}k'_{1,j}r'_{0,k}\right]^{2}\\
+\left[\frac{1}{2}\left\Vert \vec{n'}_{p,1}\right\Vert ^{-1}\epsilon_{2mn}k'_{1,m}r'_{0,n}-\grave{\nu}'_{2,2}\right]^{2}\\
\left.-\left[\frac{1}{2}\left\Vert \vec{n'}_{p,1}\right\Vert ^{-1}\epsilon_{2mn}k'_{1,m}r'_{0,n}\right]^{2}\right\} ^{1/2}\\
\times\left\{ \left[\frac{1}{2}\left\Vert \vec{n'}_{p,1}\right\Vert ^{-1}\epsilon_{3qr}k'_{1,q}r'_{0,r}-\grave{\nu}'_{2,3}\right]^{2}\right.\\
\left.\left.-\left[\frac{1}{2}\left\Vert \vec{n'}_{p,1}\right\Vert ^{-1}\epsilon_{3qr}k'_{1,q}r'_{0,r}\right]^{2}\right\} ^{-1/2}\right);\end{array}\right.\label{eq:theta2-prime-0-to-90-degree-final}\end{equation}
 \begin{equation}
\left\{ \begin{array}{c}
\lim_{\varepsilon\rightarrow0}\left(\frac{1}{2}\pi+\left|\varepsilon\right|\leq\theta'_{2}\leq\pi\right),\\
\\\theta'_{2}=\arctan\left(-\left\{ \left[\frac{1}{2}\left\Vert \vec{n'}_{p,1}\right\Vert ^{-1}\epsilon_{1jk}k'_{1,j}r'_{0,k}-\grave{\nu}'_{2,1}\right]^{2}\right.\right.\\
-\left[\frac{1}{2}\left\Vert \vec{n'}_{p,1}\right\Vert ^{-1}\epsilon_{1jk}k'_{1,j}r'_{0,k}\right]^{2}\\
+\left[\frac{1}{2}\left\Vert \vec{n'}_{p,1}\right\Vert ^{-1}\epsilon_{2mn}k'_{1,m}r'_{0,n}-\grave{\nu}'_{2,2}\right]^{2}\\
\left.-\left[\frac{1}{2}\left\Vert \vec{n'}_{p,1}\right\Vert ^{-1}\epsilon_{2mn}k'_{1,m}r'_{0,n}\right]^{2}\right\} ^{1/2}\\
\times\left\{ \left[\frac{1}{2}\left\Vert \vec{n'}_{p,1}\right\Vert ^{-1}\epsilon_{3qr}k'_{1,q}r'_{0,r}-\grave{\nu}'_{2,3}\right]^{2}\right.\\
\left.\left.-\left[\frac{1}{2}\left\Vert \vec{n'}_{p,1}\right\Vert ^{-1}\epsilon_{3qr}k'_{1,q}r'_{0,r}\right]^{2}\right\} ^{-1/2}\right),\end{array}\right.\label{eq:theta2-prime-90-to-180-degree-final}\end{equation}
 where the variables have the definition: \[
\left\{ \begin{array}{c}
\nu'_{1,1}=x'_{1}=r'_{i}\sin\theta'_{1}\cos\phi'_{1},\\
\\\nu'_{1,2}=y'_{1}=r'_{i}\sin\theta'_{1}\sin\phi'_{1},\\
\\\nu'_{1,3}=z'_{1}=r'_{i}\cos\theta'_{1};\qquad\;\:\end{array}\right.\]
\begin{align}
\theta_{inc} & =\arccos\left(\left\{ \sin\theta'_{1}\left[k'_{x'_{1}}\cos\phi'_{1}+k'_{y'_{1}}\sin\phi'_{1}\right]\right.\right.\nonumber \\
 & \left.\left.+k'_{z'_{1}}\cos\theta'_{1}\right\} \left\{ \left[k'_{x'_{1}}\right]^{2}+\left[k'_{y'_{1}}\right]^{2}+\left[k'_{z'_{1}}\right]^{2}\right\} ^{-1/2}\right),\label{eq:angle-of-incidence-exp}\end{align}
 where $k'_{1,1}=k'_{x'_{1}},$ $k'_{1,2}=k'_{y'_{1}}$ and $k'_{1,3}=k'_{z'_{1}};$
and \begin{align*}
\Gamma_{1,2} & =\left[r'_{i}\right]^{2}\sin\left(\pi-2\theta_{inc}\right),\end{align*}
 \[
\left\{ \begin{array}{ccc}
\alpha_{1}=\nu'_{1,2}+\nu'_{1,3}, &  & \alpha_{2}=\nu'_{1,3}-\nu'_{1,2},\\
\\\alpha_{3}=\nu'_{1,1}-\nu'_{1,3}, &  & \alpha_{4}=\nu'_{1,3}+\nu'_{1,1},\\
\\\alpha_{5}=\nu'_{1,1}+\nu'_{1,2}, &  & \alpha_{6}=\nu'_{1,2}-\nu'_{1,1},\end{array}\right.\]
 \[
\left\{ \begin{array}{c}
\zeta_{1}=\Gamma_{1,2}\left\Vert \vec{n'}_{p,1}\right\Vert ^{-1}\epsilon_{1jk}k'_{1,j}r'_{0,k}-\frac{1}{2}\frac{d\Gamma_{1,2}}{d\theta_{inc}},\\
\\\zeta_{2}=\Gamma_{1,2}\left\Vert \vec{n'}_{p,1}\right\Vert ^{-1}\epsilon_{2jk}k'_{1,j}r'_{0,k}-\frac{1}{2}\frac{d\Gamma_{1,2}}{d\theta_{inc}},\\
\\\zeta_{3}=\Gamma_{1,2}\left\Vert \vec{n'}_{p,1}\right\Vert ^{-1}\epsilon_{3jk}k'_{1,j}r'_{0,k}-\frac{1}{2}\frac{d\Gamma_{1,2}}{d\theta_{inc}},\end{array}\right.\]
 \begin{eqnarray*}
\widetilde{M}_{1}=\left[\begin{array}{ccc}
\zeta_{1} & \alpha_{1} & \alpha_{2}\\
\zeta_{2} & \nu'_{1,2} & \alpha_{4}\\
\zeta_{3} & \alpha_{6} & \nu'_{1,3}\end{array}\right], &  & \widetilde{M}_{2}=\left[\begin{array}{ccc}
\nu'_{1,1} & \zeta_{1} & \alpha_{2}\\
\alpha_{3} & \zeta_{2} & \alpha_{4}\\
\alpha_{5} & \zeta_{3} & \nu'_{1,3}\end{array}\right],\end{eqnarray*}
\begin{align*}
\widetilde{M}_{3} & =\left[\begin{array}{ccc}
\nu'_{1,1} & \alpha_{1} & \zeta_{1}\\
\alpha_{3} & \nu'_{1,2} & \zeta_{2}\\
\alpha_{5} & \alpha_{6} & \zeta_{3}\end{array}\right];\end{align*}
 \begin{align}
\grave{\nu}'_{2,1} & =\frac{\left[r'_{i}\right]^{-2}\det\left(\widetilde{M}_{1}\right)}{\nu'_{1,1}+\nu'_{1,2}+\nu'_{1,3}},\label{eq:x2-prime}\end{align}
\begin{align}
\grave{\nu}'_{2,2} & =\frac{\left[r'_{i}\right]^{-2}\det\left(\widetilde{M}_{2}\right)}{\nu'_{1,1}+\nu'_{1,2}+\nu'_{1,3}},\label{eq:y2-prime}\end{align}
 \begin{align}
\grave{\nu}'_{2,3} & =\frac{\left[r'_{i}\right]^{-2}\det\left(\widetilde{M}_{3}\right)}{\nu'_{1,1}+\nu'_{1,2}+\nu'_{1,3}}.\label{eq:z2-prime}\end{align}
 It can be shown that $N$th reflection point inside the hemisphere
(see Appendix A of reference \cite{key-Sung-Nae-Cho}) is, \begin{align}
\vec{R'}_{N}\left(r'_{i},\theta'_{N},\phi'_{N}\right) & =\sum_{i=1}^{3}\nu'_{N,i}\left(r'_{i},\theta'_{N},\phi'_{N}\right)\hat{e_{i}},\label{eq:Nth-Reflection-Point-NA}\end{align}
 where \[
i=\left\{ \begin{array}{c}
1\rightarrow\nu'_{N,1}=r'_{i}\sin\theta'_{N}\cos\phi'_{N},\\
\\2\rightarrow\nu'_{N,2}=r'_{i}\sin\theta'_{N}\sin\phi'_{N},\\
\\3\rightarrow\nu'_{N,3}=r'_{i}\cos\theta'_{N}.\qquad\;\:\:\end{array}\right.\]
The angular variables $\theta'_{N}$ and $\phi'_{N}$ corresponding
to $N$th reflection point $\vec{R'}_{N}$ are given (see Appendix
A of reference \cite{key-Sung-Nae-Cho}) as \begin{equation}
\left\{ \begin{array}{c}
\begin{array}{ccc}
\lim_{\varepsilon\rightarrow0}\left(0\leq\theta'_{N}\leq\frac{1}{2}\pi-\left|\varepsilon\right|\right), &  & N\geq2,\end{array}\\
\\\theta'_{N\geq2}=\arctan\left(\left\{ \left[\frac{1}{2}\left\Vert \vec{n'}_{p,1}\right\Vert ^{-1}\epsilon_{1jk}k'_{1,j}r'_{0,k}-\grave{\nu}'_{N,1}\right]^{2}\right.\right.\\
-\left[\frac{1}{2}\left\Vert \vec{n'}_{p,1}\right\Vert ^{-1}\epsilon_{1jk}k'_{1,j}r'_{0,k}\right]^{2}\\
+\left[\frac{1}{2}\left\Vert \vec{n'}_{p,1}\right\Vert ^{-1}\epsilon_{2mn}k'_{1,m}r'_{0,n}-\grave{\nu}'_{N,2}\right]^{2}\\
\left.-\left[\frac{1}{2}\left\Vert \vec{n'}_{p,1}\right\Vert ^{-1}\epsilon_{2mn}k'_{1,m}r'_{0,n}\right]^{2}\right\} ^{1/2}\\
\times\left\{ \left[\frac{1}{2}\left\Vert \vec{n'}_{p,1}\right\Vert ^{-1}\epsilon_{3qr}k'_{1,q}r'_{0,r}-\grave{\nu}'_{N,3}\right]^{2}\right.\\
\left.\left.-\left[\frac{1}{2}\left\Vert \vec{n'}_{p,1}\right\Vert ^{-1}\epsilon_{3qr}k'_{1,q}r'_{0,r}\right]^{2}\right\} ^{-1/2}\right);\end{array}\right.\label{eq:thetaN-prime-0-to-90-degree-final}\end{equation}
 \begin{equation}
\left\{ \begin{array}{c}
\begin{array}{ccc}
\lim_{\varepsilon\rightarrow0}\left(\frac{1}{2}\pi+\left|\varepsilon\right|\leq\theta'_{N}\leq\pi\right), &  & N\geq2,\end{array}\\
\\\theta'_{N\geq2}=\arctan\left(-\left\{ \left[\frac{1}{2}\left\Vert \vec{n'}_{p,1}\right\Vert ^{-1}\epsilon_{1jk}k'_{1,j}r'_{0,k}-\grave{\nu}'_{N,1}\right]^{2}\right.\right.\\
-\left[\frac{1}{2}\left\Vert \vec{n'}_{p,1}\right\Vert ^{-1}\epsilon_{1jk}k'_{1,j}r'_{0,k}\right]^{2}\\
+\left[\frac{1}{2}\left\Vert \vec{n'}_{p,1}\right\Vert ^{-1}\epsilon_{2mn}k'_{1,m}r'_{0,n}-\grave{\nu}'_{N,2}\right]^{2}\\
\left.-\left[\frac{1}{2}\left\Vert \vec{n'}_{p,1}\right\Vert ^{-1}\epsilon_{2mn}k'_{1,m}r'_{0,n}\right]^{2}\right\} ^{1/2}\\
\times\left\{ \left[\frac{1}{2}\left\Vert \vec{n'}_{p,1}\right\Vert ^{-1}\epsilon_{3qr}k'_{1,q}r'_{0,r}-\grave{\nu}'_{N,3}\right]^{2}\right.\\
\left.\left.-\left[\frac{1}{2}\left\Vert \vec{n'}_{p,1}\right\Vert ^{-1}\epsilon_{3qr}k'_{1,q}r'_{0,r}\right]^{2}\right\} ^{-1/2}\right).\end{array}\right.\label{eq:thetaN-prime-90-to-180-degree-final}\end{equation}
 \begin{equation}
\left\{ \begin{array}{c}
N\geq2,\\
\\\lim_{\varepsilon\rightarrow0}\left(0\leq\phi'_{N}\leq\frac{1}{2}\pi-\left|\varepsilon\right|\right),\\
\\\lim_{\varepsilon\rightarrow0}\left(\pi+\left|\varepsilon\right|\leq\phi'_{N}\leq\frac{3}{2}\pi-\left|\varepsilon\right|\right),\\
\\\phi'_{N\geq2}=\arctan\left(\left\{ \left[\frac{1}{2}\left\Vert \vec{n'}_{p,1}\right\Vert ^{-1}\epsilon_{2mn}k'_{1,m}r'_{0,n}-\grave{\nu}'_{N,2}\right]^{2}\right.\right.\\
\left.-\left[\frac{1}{2}\left\Vert \vec{n'}_{p,1}\right\Vert ^{-1}\epsilon_{2mn}k'_{1,m}r'_{0,n}\right]^{2}\right\} ^{1/2}\\
\times\left\{ \left[\frac{1}{2}\left\Vert \vec{n'}_{p,1}\right\Vert ^{-1}\epsilon_{1jk}k'_{1,j}r'_{0,k}-\grave{\nu}'_{N,1}\right]^{2}\right.\\
\left.\left.-\left[\frac{1}{2}\left\Vert \vec{n'}_{p,1}\right\Vert ^{-1}\epsilon_{1jk}k'_{1,j}r'_{0,k}\right]^{2}\right\} ^{-1/2}\right);\end{array}\right.\label{eq:phiN-prime-0-90-and-180-270-final}\end{equation}
 \begin{equation}
\left\{ \begin{array}{c}
N\geq2,\\
\\\lim_{\varepsilon\rightarrow0}\left(\frac{1}{2}\pi+\left|\varepsilon\right|\leq\phi'_{N}\leq\pi-\varepsilon\right),\\
\\\lim_{\varepsilon\rightarrow0}\left(\frac{3}{2}\pi+\left|\varepsilon\right|\leq\phi'_{N}<2\pi-\left|\varepsilon\right|\right),\\
\\\phi'_{N\geq2}=\arctan\left(-\left\{ \left[\frac{1}{2}\left\Vert \vec{n'}_{p,1}\right\Vert ^{-1}\epsilon_{2mn}k'_{1,m}r'_{0,n}-\grave{\nu}'_{N,2}\right]^{2}\right.\right.\\
\left.-\left[\frac{1}{2}\left\Vert \vec{n'}_{p,1}\right\Vert ^{-1}\epsilon_{2mn}k'_{1,m}r'_{0,n}\right]^{2}\right\} ^{1/2}\\
\times\left\{ \left[\frac{1}{2}\left\Vert \vec{n'}_{p,1}\right\Vert ^{-1}\epsilon_{1jk}k'_{1,j}r'_{0,k}-\grave{\nu}'_{N,1}\right]^{2}\right.\\
\left.\left.-\left[\frac{1}{2}\left\Vert \vec{n'}_{p,1}\right\Vert ^{-1}\epsilon_{1jk}k'_{1,j}r'_{0,k}\right]^{2}\right\} ^{-1/2}\right);\end{array}\right.\label{eq:phiN-prime-90-180-and-270-360-final}\end{equation}
 where $\left\{ \grave{\nu}'_{N,i}:i=1,2,3\right\} $ are found by
simply replacing in equations (\ref{eq:x2-prime}), (\ref{eq:y2-prime})
and (\ref{eq:z2-prime}) the following: \[
\left\{ \begin{array}{c}
\grave{\nu}'_{2,i}\rightarrow\grave{\nu}'_{N,i},\\
\\\zeta_{1}\left(\Gamma_{1,2}\right)\rightarrow\zeta_{1}\left(\Gamma_{1,N}\right),\\
\\\zeta_{2}\left(\Gamma_{1,2}\right)\rightarrow\zeta_{2}\left(\Gamma_{1,N}\right),\\
\\\zeta_{3}\left(\Gamma_{1,2}\right)\rightarrow\zeta_{3}\left(\Gamma_{1,N}\right),\end{array}\right.\]
 where $\Gamma_{1,N}$ is \begin{align*}
\Gamma_{1,N} & =\left[r'_{i}\right]^{2}\sin\left(\left[N-1\right]\left[\pi-2\theta_{inc}\right]\right).\end{align*}
 The details of all the work shown up to this point can be found in
Appendix A of reference \cite{key-Sung-Nae-Cho}. 

\begin{figure}
\begin{center}\includegraphics[%
  scale=0.5]{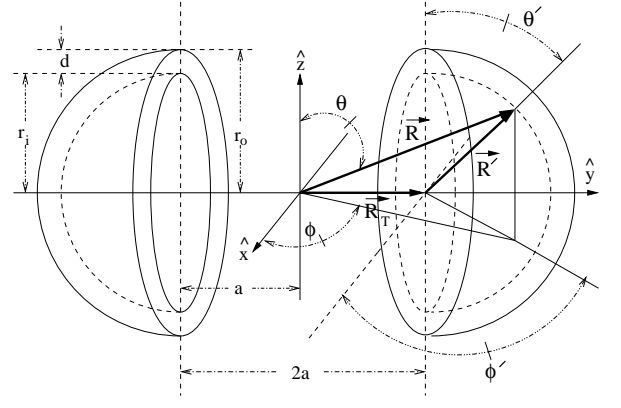}\end{center}

\caption{The surface of the hemisphere-hemisphere configuration can be described
relative to the system origin through $\vec{R},$ or relative to the
hemisphere centers through $\vec{R'}.$ \label{cap:Two-semispheres-without-center-solid-sphere}}
\end{figure}

The previously shown reflection points ($\vec{R'}_{1},$ $\vec{R'}_{2}$
and $\vec{R'}_{N}$) were described relative to the hemisphere center.
In many cases, the preferred choice for the system origin, from which
the variables are defined, depend on the physical arrangements of
the system being considered. For a sphere, the natural choice for
the origin is its center from which the spherical variables $\left(r'_{i},\theta',\phi'\right)$
are prescribed. For more complicated configuration shown in Figure
\ref{cap:Two-semispheres-without-center-solid-sphere}, the preferred
choice for origin really depends on the problem at hand. For this
reason, a set of transformation rules between $\left(r'_{i},\theta',\phi'\right)$
and $\left(r_{i},\theta,\phi\right)$ is sought. Here the primed set
is defined relative to the sphere center and the unprimed set is defined
relative to the origin of the global configuration. In terms of the
Cartesian variables, the two vectors $\vec{R}$ and $\vec{R'}$ describing
an identical point on the hemisphere surface are expressed by \begin{align}
 & \left\{ \begin{array}{c}
\vec{R}\left(\nu_{1},\nu_{2},\nu_{3}\right)=\sum_{i=1}^{3}\nu_{i}\hat{e_{i}},\\
\\\vec{R'}\left(\nu'_{1},\nu'_{2},\nu'_{3}\right)=\sum_{i=1}^{3}\nu'_{i}\hat{e_{i}},\end{array}\right.\label{eq:R-and-Rprimed-vector-cartesian-NA}\end{align}
 where \[
\left\{ \begin{array}{c}
\left(\nu_{1},\nu_{2},\nu_{3}\right)\rightarrow\left(x,y,z\right),\quad\\
\\\left(\nu'_{1},\nu'_{2},\nu'_{3}\right)\rightarrow\left(x',y',z'\right),\,\,\\
\\\left(\hat{e_{1}},\hat{e_{2}},\hat{e_{3}}\right)\rightarrow\left(\hat{x},\hat{y},\hat{z}\right).\quad\end{array}\right.\]
 The vectors $\vec{R}$ and $\vec{R'}$ are connected through the
relation \begin{align*}
\vec{R}\left(\nu_{1},\nu_{2},\nu_{3}\right) & =\sum_{i=1}^{3}\left[\nu_{T,i}+\nu'_{i}\right]\hat{e_{i}}\end{align*}
 with \begin{align*}
\vec{R}_{T} & \equiv\sum_{i=1}^{3}\nu_{T,i}\hat{e_{i}}\end{align*}
 representing the position of hemisphere center relative to the system
origin. One has then the condition \begin{align*}
\sum_{i=1}^{3}\left[\nu_{i}-\nu_{T,i}-\nu'_{i}\right]\hat{e_{i}} & =0.\end{align*}
 In terms of spherical coordinate representation for $\left(\nu_{1},\nu_{2},\nu_{3}\right)$
and $\left(\nu'_{1},\nu'_{2},\nu'_{3}\right),$ we can solve for $\theta$
and $\phi$ to yield the results (see Appendix B of reference \cite{key-Sung-Nae-Cho}):
\begin{align}
\phi & \equiv\grave{\phi}\left(r'_{i},\theta',\phi',\nu_{T,1},\nu_{T,2}\right)\nonumber \\
 & =\arctan\left(\frac{\nu_{T,2}+r'_{i}\sin\theta'\sin\phi'}{\nu_{T,1}+r'_{i}\sin\theta'\cos\phi'}\right),\label{eq:phi-NA}\end{align}
 \begin{align}
\theta & \equiv\grave{\theta}\left(r'_{i},\theta',\phi',\vec{R}_{T}\right)\nonumber \\
 & =\arctan\left(\left\{ \frac{\nu_{T,1}+\nu_{T,2}+r'_{i}\sin\theta'\left[\cos\phi'+\sin\phi'\right]}{\nu_{T,3}+r'_{i}\cos\theta'}\right\} \right.\nonumber \\
 & \times\left\{ \cos\left(\arctan\left(\frac{\nu_{T,2}+r'_{i}\sin\theta'\sin\phi'}{\nu_{T,1}+r'_{i}\sin\theta'\cos\phi'}\right)\right)\right.\nonumber \\
 & \left.\left.+\sin\left(\arctan\left(\frac{\nu_{T,2}+r'_{i}\sin\theta'\sin\phi'}{\nu_{T,1}+r'_{i}\sin\theta'\cos\phi'}\right)\right)\right\} ^{-1}\right),\label{eq:theta-NA}\end{align}
 where the notation $\grave{\phi}$ and $\grave{\theta}$ indicates
that $\phi$ and $\theta$ are explicitly expressed in terms of the
primed variables, respectively. It is to be noticed that for the configuration
shown in Figure \ref{cap:Two-semispheres-without-center-solid-sphere},
the hemisphere center is only shifted along $\hat{y}$ by an amount
of $\nu_{T,2}=a,$ which leads to $\nu_{T,i\neq2}=0.$ Nevertheless,
the derivation have been done for the case where $\nu_{T,i}\neq0,$
$i=1,2,3$ for the purpose of generalization. With the magnitude $\left\Vert \vec{R}\right\Vert $
defined as \begin{align*}
\left\Vert \vec{R}\right\Vert  & =\sqrt{\sum_{i=1}^{3}\left[\nu_{T,i}+r'_{i}\Lambda'_{i}\right]^{2}},\end{align*}
 where \[
\left\{ \begin{array}{c}
\Lambda'_{1}\left(\theta',\phi'\right)=\sin\theta'\cos\phi',\\
\\\Lambda'_{2}\left(\theta',\phi'\right)=\sin\theta'\sin\phi',\\
\\\Lambda'_{3}\left(\theta'\right)=\cos\theta',\;\;\;\;\end{array}\right.\]
 the vector $\vec{R}\left(r'_{i},\grave{\vec{\Lambda}},\vec{\Lambda}',\vec{R}_{T}\right)$
is written as \begin{align}
\vec{R}\left(r'_{i},\grave{\vec{\Lambda}},\vec{\Lambda}',\vec{R}_{T}\right) & =\sqrt{\sum_{i=1}^{3}\left[\nu_{T,i}+r'_{i}\Lambda'_{i}\right]^{2}}\sum_{i=1}^{3}\grave{\Lambda}_{i}\hat{e_{i}}.\label{eq:Points-on-Hemisphere-R-arbi-NA}\end{align}
 Here $\grave{\Lambda}_{i},$ $i=1,2,3$ is defined as \[
\left\{ \begin{array}{c}
\grave{\Lambda}_{1}\left(\grave{\theta},\grave{\phi}\right)=\sin\grave{\theta}\cos\grave{\phi},\\
\\\grave{\Lambda}_{2}\left(\grave{\theta},\grave{\phi}\right)=\sin\grave{\theta}\sin\grave{\phi}\\
\\\grave{\Lambda}_{3}\left(\grave{\theta}\right)=\cos\grave{\theta}.\;\;\;\;\end{array}\right.\]
 The details of this section can be found in Appendices A and B of
reference \cite{key-Sung-Nae-Cho}.

\subsection{Selected Configurations}

Having found all of the wave reflection points in the hemisphere resonator,
the net momentum imparted on both the inner and outer surfaces by
the incident wave is computed for three configurations: (1) the sphere,
(2) the hemisphere-hemisphere and (3) the plate-hemisphere. The surface
element that is being impinged upon by an incident wave would experience
the net momentum change in an amount proportional to $\triangle\vec{k'}_{inner}\left(;\vec{R'}_{s,1},\vec{R'}_{s,0}\right)$
on the inner side, and $\triangle\vec{k'}_{outer}\left(;\vec{R'}_{s,1}+a\hat{R'}_{s,1}\right)$
on the outer side of the surface. The quantities $\triangle\vec{k'}_{inner}$
and $\triangle\vec{k'}_{outer}$ are due to the contribution from
a single mode of wave traveling in particular direction. The notation
$\left(;\vec{R'}_{s,1},\vec{R'}_{s,0}\right)$ of $\triangle\vec{k'}_{inner}$
denotes that it is defined in terms of the initial reflection point
$\vec{R'}_{s,1}$ on the surface and the initial crossing point $\vec{R'}_{s,0}$
of the hemisphere opening (or the sphere cross-section). The notation
$\left(;\vec{R'}_{s,1}+a\hat{R'}_{s,1}\right)$ of $\triangle\vec{k'}_{outer}$
implies the outer surface reflection point. The total resultant imparted
momentum on the hemisphere or sphere is found by summing over all
modes of wave, over all directions.

\subsubsection{Hollow Spherical Shell}

\begin{figure}
\begin{center}\includegraphics[%
  scale=0.5]{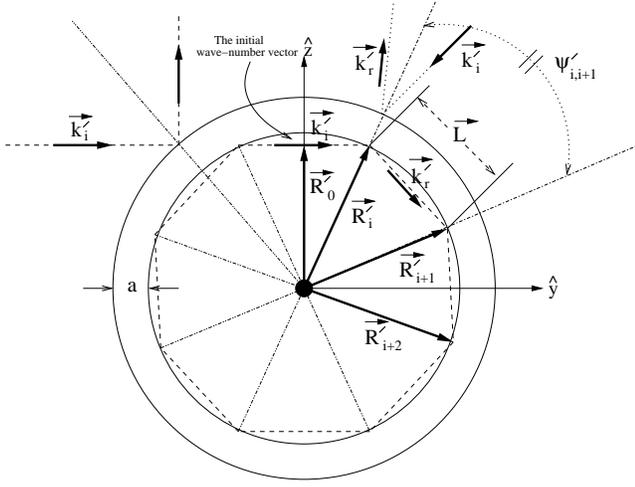}\end{center}

\caption{Inside the cavity, an incident wave $\vec{k'}_{i}$ on first impact
point $\vec{R'}_{i}$ induces a series of reflections that propagate
throughout the entire inner cavity. Similarly, a wave $\vec{k'}_{i}$
incident on the impact point $\vec{R'}_{i}+a\hat{R'}_{i},$ where
$a$ is the thickness of the sphere, induces reflected wave of magnitude
$\left\Vert \vec{k'}_{i}\right\Vert .$ The resultant wave direction
in the external region is along $\vec{R'}_{i}$ and the resultant
wave direction in the resonator is along $-\vec{R'}_{i}$ due to the
fact there is exactly another wave vector traveling in opposite direction
in both regions. In both cases, the reflected and incident waves have
equal magnitude due to the fact that the sphere is assumed to be a
perfect conductor. \label{cap:sphere-reflection-dynamics}}
\end{figure}

A sphere formed by bringing in two hemispheres together is shown in
Figure \ref{cap:sphere-reflection-dynamics}. %
 The resultant change in wave vector direction upon reflection at
the inner surface of the sphere is found to be (see Appendix C1 of
reference \cite{key-Sung-Nae-Cho}), \begin{align}
 & \triangle\vec{k'}_{inner}\left(;\vec{R'}_{s,1},\vec{R'}_{s,0}\right)\nonumber \\
 & =-\frac{4n\pi\cos\theta_{inc}}{\left\Vert \vec{R}_{s,2}\left(r'_{i},\vec{\Lambda}'_{s,2}\right)-\vec{R}_{s,1}\left(r'_{i},\vec{\Lambda}'_{s,1}\right)\right\Vert }\hat{R'}_{s,1},\label{eq:SPHERE-Delta-K-Inside-NA}\end{align}
 where \[
\left\{ \begin{array}{c}
0\leq\theta_{inc}<\pi/2,\\
\\n=1,2,\cdots.\end{array}\right.\]
 Here $\theta_{inc}$ is from equation (\ref{eq:angle-of-incidence-exp})
and, the two vectors $\vec{R}_{s,1}\left(r'_{i},\vec{\Lambda}'_{s,1}\right)$
and $\vec{R}_{s,2}\left(r'_{i},\vec{\Lambda}'_{s,2}\right)$ have
the generic form \begin{align}
\vec{R}_{s,N}\left(r'_{i},\vec{\Lambda}'_{s,N}\right) & =r'_{i}\sum_{i=1}^{3}\Lambda'_{s,N,i}\hat{e_{i}},\label{eq:Nth-Reflection-Point-sphere-NA}\end{align}
 where \[
\left\{ \begin{array}{c}
\Lambda'_{s,N,1}\left(\theta'_{s,N},\phi'_{s,N}\right)=\sin\theta'_{s,N}\cos\phi'_{s,N},\\
\\\Lambda'_{s,N,2}\left(\theta'_{s,N},\phi'_{s,N}\right)=\sin\theta'_{s,N}\sin\phi'_{s,N},\\
\\\Lambda'_{s,N,3}\left(\theta'_{s,N}\right)=\cos\theta'_{s,N}.\;\;\;\;\end{array}\right.\]
 The label $s$ has been attached to denote the sphere, and the obvious
index changes in the spherical variables $\theta'_{s,N}$ and $\phi'_{s,N}$
are understood from the set of equations (\ref{eq:thetaN-prime-0-to-90-degree-final}),
(\ref{eq:thetaN-prime-90-to-180-degree-final}), (\ref{eq:phiN-prime-0-90-and-180-270-final})
and (\ref{eq:phiN-prime-90-180-and-270-360-final}). Similarly, the
resultant change in wave vector direction upon reflection at the outer
surface of the sphere is shown to be (see Appendix C1 of reference
\cite{key-Sung-Nae-Cho}), \begin{align}
\triangle\vec{k'}_{outer}\left(;\vec{R'}_{s,1}+a\hat{R'}_{s,1}\right) & =4\left\Vert \vec{k'}_{i,f}\right\Vert \cos\theta_{inc}\hat{R'}_{s,1},\label{eq:SPHERE-Delta-K-Outside-NA}\end{align}
 where \[
\left\{ \begin{array}{c}
0\leq\theta_{inc}<\pi/2,\\
\\n=1,2,\cdots.\end{array}\right.\]
 The details of this section can be found in Appendix C1 of reference
\cite{key-Sung-Nae-Cho}.

\subsubsection{Hemisphere-Hemisphere}

For the hemisphere, the changes in wave vector directions after the
reflection at a point $\hat{R'}_{h,1}$ inside the resonator, or after
the reflection at location $\vec{R'}_{h,1}+a\hat{R'}_{h,1}$ outside
the hemisphere, can be found from equations (\ref{eq:SPHERE-Delta-K-Inside-NA})
and (\ref{eq:SPHERE-Delta-K-Outside-NA}) with obvious subscript changes,
\begin{align}
 & \triangle\vec{k'}_{inner}\left(;\vec{R'}_{h,1},\vec{R'}_{h,0}\right)\nonumber \\
 & =-\frac{4n\pi\cos\theta_{inc}}{\left\Vert \vec{R}_{h,2}\left(r'_{i},\vec{\Lambda}'_{h,2}\right)-\vec{R}_{h,1}\left(r'_{i},\vec{\Lambda}'_{h,1}\right)\right\Vert }\hat{R'}_{h,1}\label{eq:HEMISPHERE-Delta-K-Inside-NA}\end{align}
 and \begin{align}
\triangle\vec{k'}_{outer}\left(;\vec{R'}_{h,1}+a\hat{R'}_{h,1}\right) & =4\left\Vert \vec{k'}_{i,f}\right\Vert \cos\theta_{inc}\hat{R'}_{h,1},\label{eq:HEMISPHERE-Delta-K-Outside-NA}\end{align}
 where \[
\left\{ \begin{array}{c}
0\leq\theta_{inc}<\pi/2,\\
\\n=1,2,\cdots.\end{array}\right.\]
 The reflection location $\vec{R}_{h,N}\left(r'_{i},\grave{\vec{\Lambda}}_{h,N},\vec{\Lambda}'_{h,N},\vec{R}_{T,h}\right)$
has the generic form (see Appendix C2 of reference \cite{key-Sung-Nae-Cho}),
\begin{align}
 & \vec{R}_{h,N}\left(r'_{i},\grave{\vec{\Lambda}}_{h,N},\vec{\Lambda}'_{h,N},\vec{R}_{T,h}\right)\nonumber \\
 & =\sqrt{\sum_{i=1}^{3}\left[\nu_{T,h,i}+r'_{i}\Lambda'_{h,N,i}\right]^{2}}\sum_{i=1}^{3}\grave{\Lambda}_{h,N,i}\hat{e_{i}},\label{eq:Points-on-Hemisphere-R-NA}\end{align}
 where \[
\left\{ \begin{array}{c}
\grave{\Lambda}_{h,N,1}\left(\grave{\theta}_{h,N},\grave{\phi}_{h,N}\right)=\sin\grave{\theta}_{h,N}\cos\grave{\phi}_{h,N},\\
\\\grave{\Lambda}_{h,N,2}\left(\grave{\theta}_{h,N},\grave{\phi}_{h,N}\right)=\sin\grave{\theta}_{h,N}\sin\grave{\phi}_{h,N},\\
\\\grave{\Lambda}_{h,N,3}\left(\grave{\theta}_{h,N}\right)=\cos\grave{\theta}_{h,N}.\;\;\;\;\end{array}\right.\]
 The subscript $h$ here denotes the hemisphere. The expressions for
$\Lambda'_{h,N,i},$ $i=1,2,3,$ are defined identically in form.
The angular variables in spherical coordinates, $\grave{\theta}_{h,N}$
and $\grave{\phi}_{h,N},$ can be obtained from equations (\ref{eq:phi-NA})
and (\ref{eq:theta-NA}), where the obvious notational changes are
understood. The implicit angular variables, $\theta'_{h,N}$ and $\phi'_{h,N},$
are the sets defined in equations (\ref{eq:thetaN-prime-0-to-90-degree-final})
and (\ref{eq:thetaN-prime-90-to-180-degree-final}) for $\theta'_{s,N},$
and the sets from equations (\ref{eq:phiN-prime-0-90-and-180-270-final})
and (\ref{eq:phiN-prime-90-180-and-270-360-final}) for $\phi'_{s,N}.$ 

Unlike the sphere situation, the initial wave could eventually escape
the hemisphere resonator after some maximum number of reflections.
It is shown (see Appendix C2 of reference\cite{key-Sung-Nae-Cho})
that this maximum number for internal reflection is given by \begin{align}
N_{h,max} & =\left[\mathbb{Z}_{h,max}\right]_{G},\label{eq:N-max-Hemisphere-NA}\end{align}
 where the notation $\left[\mathbb{Z}_{h,max}\right]_{G}$ denotes
the greatest integer function, and $\mathbb{Z}_{h,max}$ is given
by \begin{align}
\mathbb{Z}_{h,max} & =\frac{1}{\pi-2\theta_{inc}}\left[\pi-\arccos\left(\frac{1}{2}\left\{ r'_{i}\left\Vert \vec{R'}_{0}\right\Vert ^{-1}\right.\right.\right.\nonumber \\
 & \left.\left.\left.+\left[r'_{i}\right]^{-1}\left\Vert \vec{R'}_{0}\right\Vert -\left[r'_{i}\left\Vert \vec{R'}_{0}\right\Vert \right]^{-1}\xi_{1,p}^{2}\right\} \right)\right].\label{eq:N-max-Hemisphere-Greatest-Integer-Function-NA}\end{align}
 Here $\xi_{1,p}$ is from equation (\ref{eq:positive-root-k-NA})
and $\theta_{inc}$ is from equation (\ref{eq:angle-of-incidence-exp}). 

\begin{figure}
\begin{center}\includegraphics[%
  scale=0.5]{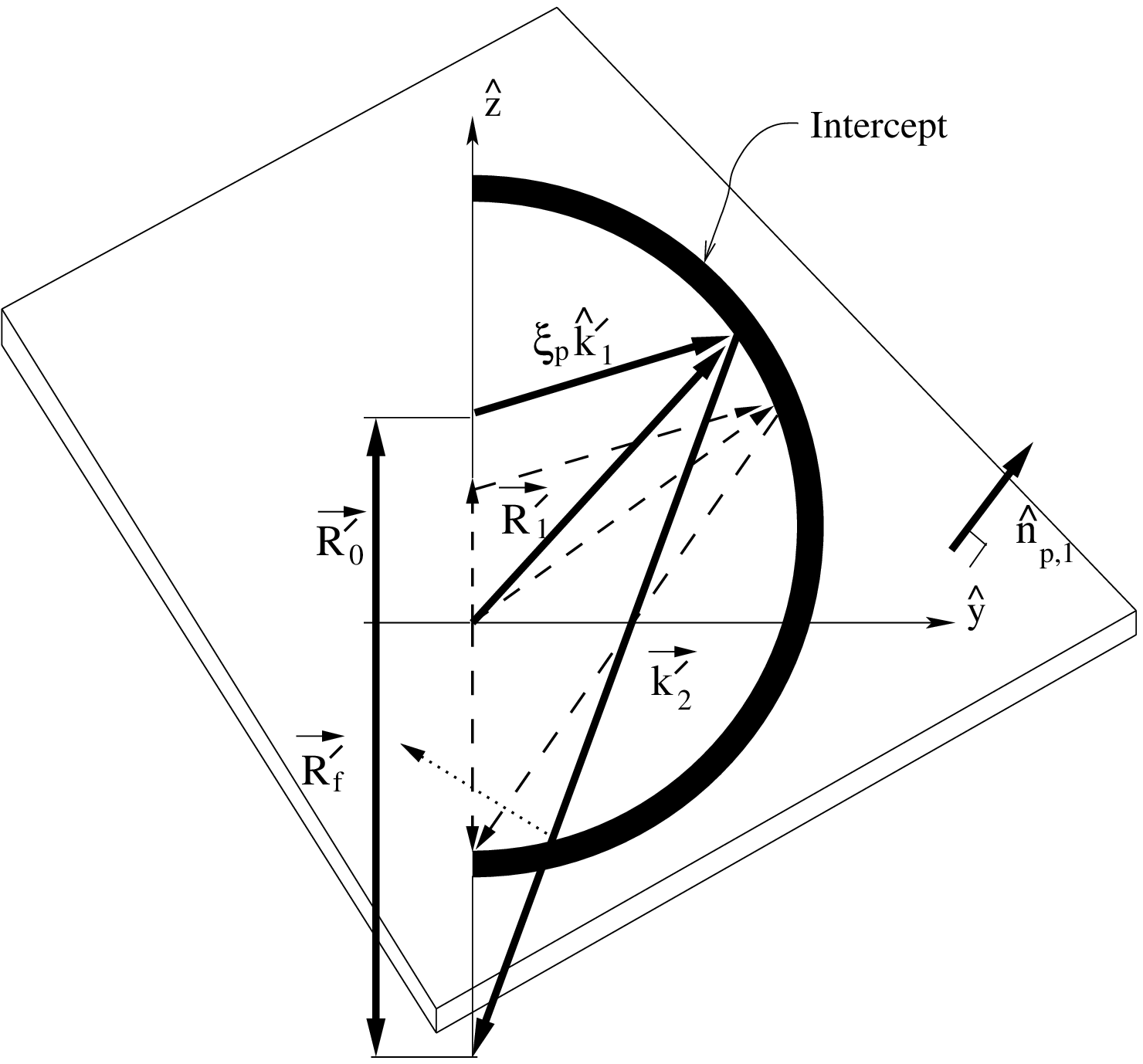}\end{center}

\caption{The dashed line vectors represent the situation where only single
internal reflection occurs. The dark line vectors represent the situation
where multiple internal reflections occur. \label{cap:plane-sphere-intersection-3-Critical-Angle}}
\end{figure}

The above results of $\triangle\vec{k'}_{inner}\left(;\vec{R'}_{h,1},\vec{R'}_{h,0}\right)$
and $\triangle\vec{k'}_{outer}\left(;\vec{R'}_{h,1}+a\hat{R'}_{h,1}\right)$
have been derived based on the fact that there are multiple internal
reflections. For a sphere, the multiple internal reflections are inherent.
However, for a hemisphere, it is not necessarily true that all incoming
waves would result in multiple internal reflections. Naturally, the
criteria for multiple internal reflections are in order. If the initial
direction of the incoming wave vector, $\hat{k'}_{1},$ is given,
the internal reflections can be either single or multiple depending
upon the location of the entry point in the cavity, $\vec{R'}_{0}.$
As shown in Figure \ref{cap:plane-sphere-intersection-3-Critical-Angle},
these are two reflection dynamics where the dashed vectors represent
the single reflection case and the non-dashed vectors represent multiple
reflections case. Because the whole process occurs in the same plane
of incidence, the vector $\vec{R'}_{f}=-\lambda_{0}\vec{R'}_{0}$
where $\lambda_{0}>0.$ The multiple or single internal reflection
criteria can be summarized by the relation (see Appendix C2 of reference
\cite{key-Sung-Nae-Cho}): \begin{align}
\left\Vert \vec{R'}_{f}\right\Vert  & =\frac{1}{2}\left\Vert \vec{R'}_{0}\right\Vert \left[\sum_{n=1}^{3}k'_{1,n}\right]\left\{ \sum_{j=1}^{3}\sum_{l=1}^{3}\left[\left\Vert \vec{k'}_{1}\right\Vert ^{2}-k'_{1,l}\right.\right.\nonumber \\
 & \left.\left.\times\sum_{m=1}^{3}k'_{1,m}\right]r'_{0,l}r'_{0,j}\right\} ^{-1}\sum_{l=1}^{3}\left\{ k'_{1,l}\left[r'_{i}\right]^{2}-\left[r'_{0,l}\right]^{2}\right.\nonumber \\
 & +2\vec{R'}_{0}\cdot\vec{k'}_{1}r'_{0,l}-\left\Vert \vec{R'}_{0}\right\Vert ^{2}k'_{1,l}-2r'_{0,l}\left[\sum_{l=1}^{3}k'_{1,l}\right]^{-1}\nonumber \\
 & \left.\times\sum_{i=1}^{3}\left[\left\Vert \vec{k'}_{1}\right\Vert ^{2}-k'_{1,i}\sum_{m=1}^{3}k'_{1,m}\right]r'_{0,i}\right\} .\label{eq:Hem-Hemi-Multi-Single-Ref-condi-NA}\end{align}
 Because the hemisphere opening has a radius $r'_{i},$ the following
criteria are concluded: \begin{equation}
\left\{ \begin{array}{ccc}
\left\Vert \vec{R'}_{f}\right\Vert <r'_{i}, &  & single\; internal\; reflection,\\
\\\left\Vert \vec{R'}_{f}\right\Vert \geq r'_{i}, &  & multiple\; internal\; reflections,\end{array}\right.\label{eq:Hem-Hemi-Multi-Single-Ref-condi-Interpretation-NA}\end{equation}
 where $\left\Vert \vec{R'}_{f}\right\Vert $ is defined in equation
(\ref{eq:Hem-Hemi-Multi-Single-Ref-condi-NA}). The details of this
section can be found in Appendix C2 of reference \cite{key-Sung-Nae-Cho}.

\subsubsection{Plate-Hemisphere}

\begin{figure}
\begin{center}\includegraphics[%
  scale=0.5]{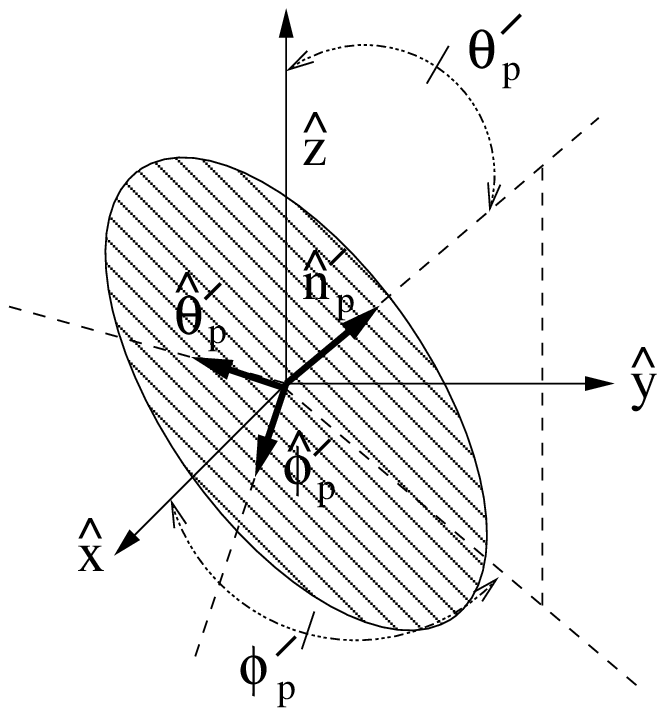}\end{center}

\caption{The orientation of a disk is given through the surface unit normal
$\hat{n'}_{p}.$ The disk is spanned by the two unit vectors $\hat{\theta'}_{p}$
and $\hat{\phi'}_{p}.$ \label{cap:circular-plane}}
\end{figure}

A surface is represented by a unit vector $\hat{n'}_{p},$ which is
normal to the surface locally. For the circular plate shown in Figure
\ref{cap:circular-plane}, its orthonormal triad $\left(\hat{n'}_{p},\hat{\theta'}_{p},\hat{\phi'}_{p}\right)$
has the form \begin{eqnarray*}
\hat{n'}_{p}=\sum_{i=1}^{3}\Lambda'_{p,i}\hat{e}_{i}, &  & \hat{\theta'}_{p}=\sum_{i=1}^{3}\frac{\partial\Lambda'_{p,i}}{\partial\theta'_{p}}\hat{e}_{i},\end{eqnarray*}
 \[
\hat{\phi'}_{p}=\sum_{i=1}^{3}\frac{1}{\sin\theta'_{p}}\frac{\partial\Lambda'_{p,i}}{\partial\phi'_{p}}\hat{e}_{i},\]
 where \[
\left\{ \begin{array}{c}
\Lambda'_{p,1}\left(\theta'_{p},\phi'_{p}\right)=\sin\theta'_{p}\cos\phi'_{p},\\
\\\Lambda'_{p,2}\left(\theta'_{p},\phi'_{p}\right)=\sin\theta'_{p}\sin\phi'_{p},\\
\\\Lambda'_{p,3}\left(\theta'_{p}\right)=\cos\theta'_{p}.\;\;\;\;\end{array}\right.\]

\begin{figure}
\begin{center}\includegraphics[%
  scale=0.5]{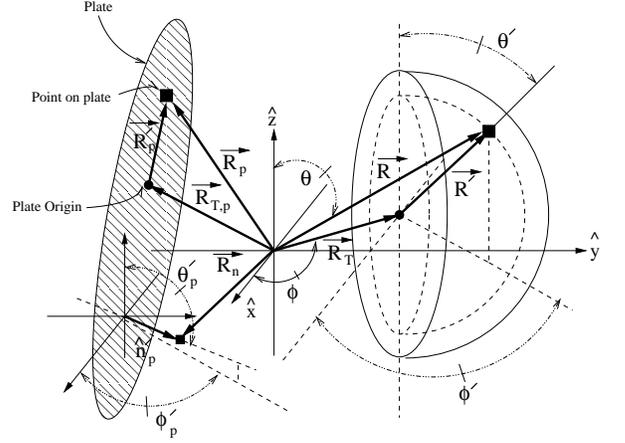}\end{center}

\caption{The plate-hemisphere configuration. \label{cap:plate-and-hemisphere-complex}}
\end{figure}

For the plate-hemisphere configuration shown in Figure \ref{cap:plate-and-hemisphere-complex},
it can be shown that the element $\vec{R}_{p}$ on the plane and its
velocity $d\vec{R}_{p}/dt$ are given by (see Appendix C3 of reference
\cite{key-Sung-Nae-Cho}): \begin{align}
 & \vec{R}_{p}\left(\grave{\vec{\Lambda}}_{p},\vec{\Lambda}'_{p},\vec{R}_{T,p}\right)\nonumber \\
 & =\left\{ \sum_{i=1}^{3}\left[\nu_{T,p,i}+\nu'_{p,\theta'_{p}}\frac{\partial\Lambda'_{p,i}}{\partial\theta'_{p}}+\frac{\nu'_{p,\phi'_{p}}}{\sin\theta'_{p}}\frac{\partial\Lambda'_{p,i}}{\partial\phi'_{p}}\right]^{2}\right\} ^{1/2}\nonumber \\
 & \times\sum_{i=1}^{3}\grave{\Lambda}_{p,i}\hat{e_{i}}\label{eq:Points-on-Plate-R-NA}\end{align}
 and \begin{align}
\dot{\vec{R}}_{p} & \equiv\frac{d\vec{R}_{p}}{dt}\nonumber \\
 & =\left\{ \sum_{i=1}^{3}\left[\nu_{T,p,i}+\nu'_{p,\theta'_{p}}\frac{\partial\Lambda'_{p,i}}{\partial\theta'_{p}}+\frac{\nu'_{p,\phi'_{p}}}{\sin\theta'_{p}}\frac{\partial\Lambda'_{p,i}}{\partial\phi'_{p}}\right]^{2}\right\} ^{-1/2}\nonumber \\
 & \times\sum_{j=1}^{3}\sum_{k=1}^{3}\left(\left[\nu_{T,p,k}+\nu'_{p,\theta'_{p}}\frac{\partial\Lambda'_{p,k}}{\partial\theta'_{p}}+\frac{\nu'_{p,\phi'_{p}}}{\sin\theta'_{p}}\frac{\partial\Lambda'_{p,k}}{\partial\phi'_{p}}\right]\right.\nonumber \\
 & \times\left[\dot{\nu}_{T,p,k}+\left\{ \nu'_{p,\theta'_{p}}\frac{\partial^{2}\Lambda'_{p,k}}{\partial\left[\theta'_{p}\right]^{2}}+\frac{\nu'_{p,\phi'_{p}}}{\sin\theta'_{p}}\left(\frac{\partial^{2}\Lambda'_{p,k}}{\partial\theta'_{p}\partial\phi'_{p}}\right.\right.\right.\nonumber \\
 & \left.\left.-\cot\theta'_{p}\frac{\partial\Lambda'_{p,k}}{\partial\phi'_{p}}\right)\right\} \dot{\theta'}_{p}+\left\{ \nu'_{p,\theta'_{p}}\frac{\partial^{2}\Lambda'_{p,k}}{\partial\phi'_{p}\partial\theta'_{p}}+\frac{\nu'_{p,\phi'_{p}}}{\sin\theta'_{p}}\right.\nonumber \\
 & \left.\left.\times\frac{\partial^{2}\Lambda'_{p,k}}{\partial\left[\phi'_{p}\right]^{2}}\right\} \dot{\phi'}_{p}+\dot{\nu'}_{p,\theta'_{p}}\frac{\partial\Lambda'_{p,k}}{\partial\theta'_{p}}+\frac{\dot{\nu'}_{p,\phi'_{p}}}{\sin\theta'_{p}}\frac{\partial\Lambda'_{p,k}}{\partial\phi'_{p}}\right]\nonumber \\
 & \times\grave{\Lambda}_{p,j}+\sum_{i=1}^{3}\left[\nu_{T,p,i}+\nu'_{p,\theta'_{p}}\frac{\partial\Lambda'_{p,i}}{\partial\theta'_{p}}+\frac{\nu'_{p,\phi'_{p}}}{\sin\theta'_{p}}\frac{\partial\Lambda'_{p,i}}{\partial\phi'_{p}}\right]^{2}\nonumber \\
 & \left.\times\left[\frac{\partial\grave{\Lambda}_{p,j}}{\partial\grave{\theta}_{p}}\frac{\partial\grave{\theta}_{p}}{\partial\phi'_{p}}\dot{\theta}'_{p}+\frac{\partial\grave{\Lambda}_{p,j}}{\partial\grave{\phi}_{p}}\frac{\partial\grave{\phi}_{p}}{\partial\phi'_{p}}\dot{\phi}'_{p}\right]\right)\hat{e_{j}},\label{eq:velocity-points-on-circular-plane-NA}\end{align}
 where \begin{align}
\grave{\phi}_{p} & \equiv\grave{\phi}_{p}\left(\theta'_{p},\phi'_{p},\nu_{T,p,1},\nu_{T,p,2}\right)\nonumber \\
 & =\arctan\left(\frac{\nu_{T,p,2}+\sin\theta'_{p}\sin\phi'_{p}}{\nu_{T,p,1}+\sin\theta'_{p}\cos\phi'_{p}}\right),\label{eq:phi-Plate}\end{align}
 \begin{align}
\grave{\theta}_{p} & \equiv\grave{\theta}_{p}\left(\theta'_{p},\phi'_{p},\vec{R}_{T,p}\right)\nonumber \\
 & =\arctan\left(\frac{\nu_{T,p,1}+\nu_{T,p,2}+\sin\theta'_{p}\left[\cos\phi'_{p}+\sin\phi'_{p}\right]}{\nu_{T,p,3}+\cos\theta'_{p}}\right.\nonumber \\
 & \times\left\{ \cos\left(\arctan\left(\frac{\nu_{T,p,2}+\sin\theta'_{p}\sin\phi'_{p}}{\nu_{T,p,1}+\sin\theta'_{p}\cos\phi'_{p}}\right)\right)\right.\nonumber \\
 & \left.\left.+\sin\left(\arctan\left(\frac{\nu_{T,p,2}+\sin\theta'_{p}\sin\phi'_{p}}{\nu_{T,p,1}+\sin\theta'_{p}\cos\phi'_{p}}\right)\right)\right\} ^{-1}\right),\label{eq:theta-Plate}\end{align}
 and \begin{equation}
\left\{ \begin{array}{c}
\grave{\Lambda}_{p,1}\left(\grave{\theta}_{p},\grave{\phi}_{p}\right)=\sin\grave{\theta}_{p}\cos\grave{\phi}_{p},\\
\\\grave{\Lambda}_{p,2}\left(\grave{\theta}_{p},\grave{\phi}_{p}\right)=\sin\grave{\theta}_{p}\sin\grave{\phi}_{p},\\
\\\grave{\Lambda}_{p,3}\left(\grave{\theta}_{p}\right)=\cos\grave{\theta}_{p}.\;\;\;\;\end{array}\right.\label{eq:Capital-Lambda-Plate-Def}\end{equation}
 The subscript $p$ of $\grave{\phi}_{p}$ and $\grave{\theta}_{p}$
indicates that these are spherical variables for the points on the
plate of Figure \ref{cap:plate-and-hemisphere-complex}, not that
of the hemisphere. It is also understood that $\Lambda'_{p,3}$ and
$\grave{\Lambda}_{p,3}$ are independent of $\phi'_{p}$ and $\grave{\phi}_{p},$
respectively. Therefore, their differentiation with respect to $\phi'_{p}$
and $\grave{\phi}_{p}$ respectively vanishes. The quantities $\dot{\theta'}_{p}$
and $\dot{\phi'}_{p}$ are the angular frequencies, and $\dot{\nu}_{T,p,i}$
is the translation speed of the plate relative to the system origin.
The quantities $\dot{\nu'}_{p,\theta'_{p}}$ and $\dot{\nu'}_{p,\phi'_{p}}$
are the lattice vibrations along the directions $\hat{\theta'}_{p}$
and $\hat{\phi'}_{p}$ respectively. For the static plate without
lattice vibrations, $\dot{\nu'}_{p,\theta'_{p}}$ and $\dot{\nu'}_{p,\phi'_{p}}$
vanishes. 

\begin{figure}[t]
\begin{center}\includegraphics[%
  scale=0.4]{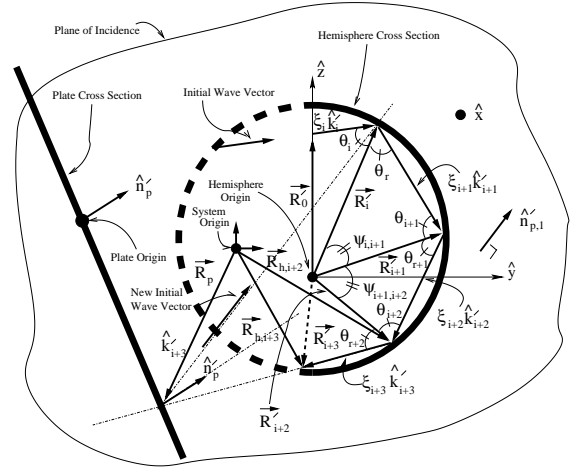}\end{center}

\caption{The intersection between oscillating plate, hemisphere and the plane
of incidence whose normal is $\hat{n'}_{p,1}=-\left\Vert \vec{n'}_{p,1}\right\Vert ^{-1}\sum_{i=1}^{3}\epsilon_{ijk}k'_{1,j}r'_{0,k}\hat{e_{i}}.$
\label{cap:plate-hemisphere-plane-of-incidence-intersect-Complex}}
\end{figure}

In the cross-sectional view of the plate-hemisphere system shown in
Figure \ref{cap:plate-hemisphere-plane-of-incidence-intersect-Complex},
the initial wave vector $\vec{k'}_{i}$ traveling toward the hemisphere
would go through a series of reflections according to the law of reflection
and finally exit the cavity. It would then continue toward the plate,
and depending on the orientation of plate at the time of impact, the
wave-vector, now reflecting off the plate, would either escape to
infinity or re-enter the hemisphere. The process repeats successively.
In order to determine whether the wave that just escaped from the
hemisphere cavity can reflect back from the plate and re-enter the
hemisphere or escape to infinity, the exact location of reflection
on the plate must be known. This reflection point on the plate is
found to be (see Appendix C3 of reference \cite{key-Sung-Nae-Cho}),
\begin{align}
\vec{R}_{p}= & \left\{ \sum_{s=1}^{3}\left[-\frac{\sum_{i=1}^{3}\frac{\partial\Lambda'_{p,i}}{\partial\phi'_{p}}\left(\Lambda'_{p,i}+\frac{\epsilon_{ijk}k'_{1,j}r'_{0,k}}{\left\Vert \vec{n'}_{p,1}\right\Vert }\right)}{\sum_{l=1}^{3}\frac{\partial\Lambda'_{p,l}}{\partial\theta'_{p}}\left(\Lambda'_{p,l}+\frac{\epsilon_{lmn}k'_{1,m}r'_{0,n}}{\left\Vert \vec{n'}_{p,1}\right\Vert }\right)}\right.\right.\nonumber \\
 & \left.\left.\times\frac{\partial\Lambda'_{p,s}}{\partial\theta'_{p}}+\frac{\partial\Lambda'_{p,s}}{\partial\phi'_{p}}\right]^{2}\right\} ^{1/2}\left[\frac{A_{\gamma}A_{\beta}}{C_{\beta}C_{\gamma}}+\frac{B_{\zeta}B_{\beta}}{C_{\beta}}\right.\nonumber \\
 & \left.+\frac{\gamma_{o}B_{\gamma}A_{\beta}}{C_{\beta}C_{\gamma}}\right]\sum_{i=1}^{3}\grave{\Lambda}_{p,i}\hat{e_{i}},\label{eq:Points-on-Plate-R-Final-NA}\end{align}
 where the translation parameter $\nu_{T,p,j}=0;$ and the terms $\left(A_{\zeta},B_{\zeta},C_{\zeta}\right),$
$\left(A_{\gamma},B_{\gamma},C_{\gamma}\right),$ $\left(A_{\beta},B_{\beta},C_{\beta}\right)$
and $\gamma_{o}$ are defined as (see Appendix C3 of reference \cite{key-Sung-Nae-Cho}):
\begin{equation}
\left\{ \begin{array}{c}
A_{\zeta}=\sqrt{\sum_{j=1}^{3}\left[\nu_{T,h,j}+r'_{i}\Lambda'_{h,N_{h,max}+1,j}\right]^{2}},\\
\\B_{\zeta}=\sqrt{\sum_{j=1}^{3}\left[\nu_{T,h,j}+r'_{i}\Lambda'_{h,N_{h,max},j}\right]^{2}},\\
\\C_{\zeta}=-\left(\sum_{x=1}^{3}\frac{\partial\Lambda'_{p,x}}{\partial\phi'_{p}}\left[\Lambda'_{p,x}+\frac{\epsilon_{xyz}k'_{1,y}r'_{0,z}}{\left\Vert \vec{n'}_{p,1}\right\Vert }\right]\right)\\
\times\left(\sum_{l=1}^{3}\frac{\partial\Lambda'_{p,l}}{\partial\theta'_{p}}\left[\Lambda'_{p,l}+\frac{\epsilon_{lmn}k'_{1,m}r'_{0,n}}{\left\Vert \vec{n'}_{p,1}\right\Vert }\right]\right)^{-1};\end{array}\right.\label{eq:Plate-Hemi-exiting-K-scale-ABC-in-Zetai-Def}\end{equation}
 \begin{equation}
\left\{ \begin{array}{c}
A_{\gamma}=\epsilon_{ijk}R_{h,N_{h,max},j}\left[R_{h,N_{h,max}+1,k}-R_{h,N_{h,max},k}\right],\\
\\B_{\gamma}=\left\Vert \vec{n'}_{p,1}\right\Vert ^{-1}\epsilon_{ijk}k'_{1,j}r'_{0,k},\\
\\C_{\gamma}=\epsilon_{ijk}\left[R_{h,N_{h,max},j}-R_{h,N_{h,max}+1,j}\right]\\
\times\left[R_{h,N_{h,max}+1,k}-R_{h,N_{h,max},k}\right];\end{array}\right.\label{eq:Plate-Hemi-exiting-K-scale-ABC-in-GAMMA-Def}\end{equation}
 \begin{equation}
\left\{ \begin{array}{c}
A_{\beta}=\sum_{i=1}^{3}\left[A_{\zeta}\grave{\Lambda}_{h,N_{h,max}+1,i}-B_{\zeta}\grave{\Lambda}_{h,N_{h,max},i}\right],\\
\\B_{\beta}=\sum_{i=1}^{3}\grave{\Lambda}_{h,N_{h,max},i},\\
\\C_{\beta}=\left\{ \sum_{j=1}^{3}\left[\frac{\partial\Lambda'_{p,j}}{\partial\phi'_{p}}-C_{\zeta}\frac{\partial\Lambda'_{p,j}}{\partial\theta'_{p}}\right]^{2}\right\} ^{1/2}\\
\times\sum_{l=1}^{3}\grave{\Lambda}_{p,l};\end{array}\right.\label{eq:Plate-Hemi-exiting-K-scale-ABC-in-BETA-Def}\end{equation}
 \begin{align}
\gamma_{o} & =\left(\left\Vert \vec{n'}_{p,1}\right\Vert ^{-1}\sum_{i=1}^{3}\epsilon_{ijk}k'_{1,j}r'_{0,k}-C_{\beta}^{-1}C_{\gamma}^{-1}\right.\nonumber \\
 & \times B_{\gamma}A_{\beta}\sqrt{\sum_{j=1}^{3}\left[\frac{\partial\Lambda'_{p,j}}{\partial\phi'_{p}}-C_{\zeta}\frac{\partial\Lambda'_{p,j}}{\partial\theta'_{p}}\right]^{2}}\nonumber \\
 & \left.\times\sum_{i=1}^{3}\epsilon_{ijk}\grave{\Lambda}_{p,j}k_{N_{h,max}+1,k}\right)^{-1}\left(\left[\frac{A_{\gamma}A_{\beta}}{C_{\beta}C_{\gamma}}\right.\right.\nonumber \\
 & \left.+\frac{B_{\zeta}B_{\beta}}{C_{\beta}}\right]\sqrt{\sum_{j=1}^{3}\left[\frac{\partial\Lambda'_{p,j}}{\partial\phi'_{p}}-C_{\zeta}\frac{\partial\Lambda'_{p,j}}{\partial\theta'_{p}}\right]^{2}}\nonumber \\
 & \left.\times\sum_{i=1}^{3}\epsilon_{ijk}\grave{\Lambda}_{p,j}k_{N_{h,max}+1,k}\right).\label{eq:Plate-Hemi-exiting-K-scale-GAMMA}\end{align}
It is to be noticed that for a situation where the translation parameter
$\nu_{T,p,j}=0,$ the $\grave{\Lambda}$ becomes identical to $\Lambda'$
in form. Results for $\grave{\Lambda}$ can be obtained from $\Lambda'$
by a simple replacement of primed variables with the unprimed ones.
It can be shown that the criterion whether the wave reflecting off
the plate at location $\vec{R}_{p}$ can re-enter the hemisphere cavity
or simply escape to infinity is found from the relation (see Appendix
C3 of reference \cite{key-Sung-Nae-Cho}), \begin{align}
\xi_{\kappa,i} & =\left(\nu_{T,h,i}+r'_{0,i}-\left\{ \sum_{s=1}^{3}\left[\frac{\partial\Lambda'_{p,s}}{\partial\phi'_{p}}-\frac{\partial\Lambda'_{p,s}}{\partial\theta'_{p}}\right.\right.\right.\nonumber \\
 & \left.\left.\times\frac{\sum_{i=1}^{3}\frac{\partial\Lambda'_{p,i}}{\partial\phi'_{p}}\left(\Lambda'_{p,i}+\frac{\epsilon_{ijk}k'_{1,j}r'_{0,k}}{\left\Vert \vec{n'}_{p,1}\right\Vert }\right)}{\sum_{l=1}^{3}\frac{\partial\Lambda'_{p,l}}{\partial\theta'_{p}}\left(\Lambda'_{p,l}+\frac{\epsilon_{lmn}k'_{1,m}r'_{0,n}}{\left\Vert \vec{n'}_{p,1}\right\Vert }\right)}\right]^{2}\right\} ^{1/2}\nonumber \\
 & \left.\times\left[\frac{A_{\gamma}A_{\beta}}{C_{\beta}C_{\gamma}}+\frac{\gamma_{o}B_{\gamma}A_{\beta}}{C_{\beta}C_{\gamma}}+\frac{B_{\zeta}B_{\beta}}{C_{\beta}}\right]\grave{\Lambda}_{p,i}\right)\nonumber \\
 & \times\left(\sum_{k=1}^{3}\left\{ \alpha_{r,\perp}\left[k_{N_{h,max}+1,i}n'_{p,k}n'_{p,k}\right.\right.\right.\nonumber \\
 & \left.-n'_{p,i}k_{N_{h,max}+1,k}n'_{p,k}\right]-\alpha_{r,\parallel}n'_{p,k}\nonumber \\
 & \left.\left.\times k_{N_{h,max}+1,k}n'_{p,i}\right\} \right)^{-1},\label{eq:Plate-Hemi-K-Reenter-Criteria-NA}\end{align}
 where $i=1,2,3$ and $\xi_{\kappa,i}$ is the component of the scale
vector $\vec{\xi}_{\kappa}=\sum_{i=1}^{3}\xi_{\kappa,i}\hat{e_{i}}$
explained in the Appendix C3 of reference \cite{key-Sung-Nae-Cho}. 

In the above re-entry criteria, it should be noticed that $\vec{R}_{0}\leq r'_{i}.$
This implies $r'_{0,i}\leq r'_{i},$ where $r'_{i}$ is the radius
of hemisphere. It is then concluded that all waves re-entering the
hemisphere cavity would satisfy the condition $\xi_{\kappa,1}=\xi_{\kappa,2}=\xi_{\kappa,3}.$
On the other hand, those waves that escapes to infinity cannot have
all three $\xi_{\kappa,i}$ equal to a single constant. The re-entry
condition $\xi_{\kappa,1}=\xi_{\kappa,2}=\xi_{\kappa,3}$ is just
another way of stating the existence of a parametric line along the
vector $\vec{k}_{r,N_{h,max}+1}$ that happens to pierce through the
hemisphere opening. In case such a line does not exist, the initial
wave direction has to be rotated into a new direction such that there
is a parametric line that pierces through the hemisphere opening.
That is why all three $\xi_{\kappa,i}$ values cannot be equal to
a single constant. The re-entry criteria are summarized here for bookkeeping
purpose: \begin{equation}
\left\{ \begin{array}{c}
\xi_{\kappa,1}=\xi_{\kappa,2}=\xi_{\kappa,3}:\quad wave\; reenters\; hemisphere,\\
\\\;\;\;\qquad\qquad else:\quad wave\; escapes\; to\; infinity,\end{array}\right.\label{eq:Plate-Hemi-K-Reenter-Criteria-Interpretation-NA}\end{equation}
 where $else$ is the case where $\xi_{\kappa,1}=\xi_{\kappa,2}=\xi_{\kappa,3}$
cannot be satisfied. The details of this section can be found in Appendix
C3 of reference \cite{key-Sung-Nae-Cho}.

\subsection{Dynamical Casimir Force}

\begin{figure}
\begin{center}\includegraphics[%
  scale=0.5]{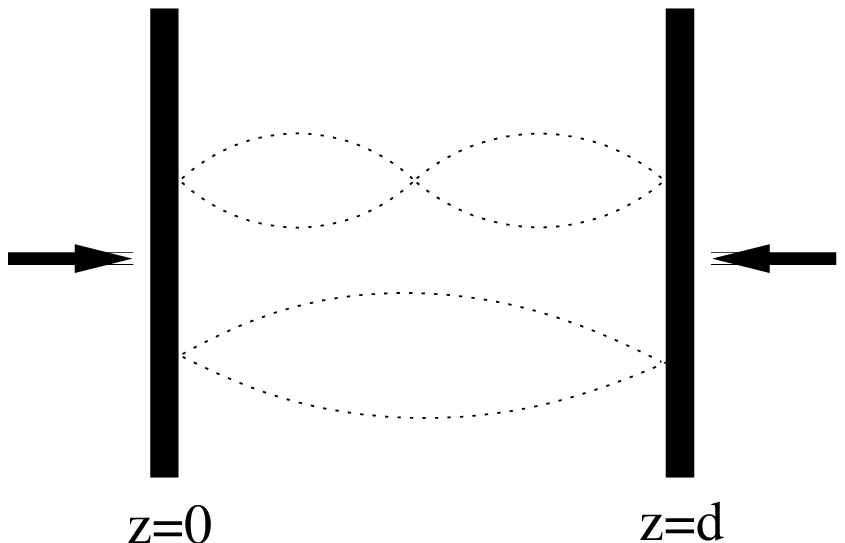}\end{center}

\caption{Because there are more vacuum-field modes in the external regions,
the two charge-neutral conducting plates are accelerated inward till
the two finally stick.\label{cap:casimir-plates-outlook}}
\end{figure}

The phenomenon of Casimir effect is inherently a dynamical effect
due to the fact that it involves radiation, rather than static fields.
One of our original objectives in studying the Casimir effect was
to investigate the physical implications of vacuum-fields on movable
boundaries. Consider the two parallel plates configuration of charge-neutral,
perfect conductors shown in Figure \ref{cap:casimir-plates-outlook}.
Because there are more wave modes in the outer region of the parallel
plate resonator, two loosely restrained (or unfixed in position) plates
will accelerate inward until they finally meet. The energy conservation
would require that the energy initially confined in the resonator
when the two plates were separated be transformed into the heat energy
that acts to raise the temperatures of the two plates. 

Davies in 1975 \cite{key-Davies}, followed by Unruh in 1976 \cite{key-Unruh},
have asked the similar question and came to a conclusion that when
an observer is moving with a constant acceleration in vacuum, the
observer perceives himself to be immersed in a thermal bath at the
temperature \begin{align*}
T & =\frac{\hbar\ddot{R}}{2\pi ck'},\end{align*}
 where $\ddot{R}$ is the acceleration of the observer and $k',$
the wave number. The details of the Unruh-Davies effect can also be
found in the reference \cite{key-Milonni}. The other work that dealt
with the concept of dynamical Casimir effect is due to Schwinger in
his proposals \cite{key-Milton,key-Schwinger-Casimir-Light} to explain
the phenomenon of sonoluminescense. Sonoluminescense is a phenomenon
in which when a small air bubble filled with noble gas is under a
strong acoustic-field pressure, the bubble will emit an intense flash
of light in the optical range. 

Our formulation of dynamical Casimir effect here, however, has no
resemblance to that of Schwinger's work to the best of our knowledge.
This formulation of dynamical Casimir force is briefly presented in
the following sections. The details of derivations pertaining to the
dynamical Casimir force can be found in Appendix D of reference \cite{key-Sung-Nae-Cho}.

\subsubsection{Formalism of Zero-Point Energy and its Force}

For massless fields, the energy-momentum relation is \begin{align}
\mathcal{H}'_{n_{s}} & \equiv E_{Total}=pc,\label{eq:2-3-1}\end{align}
 where $p$ is the momentum, $c$ is the speed of light, and $\mathcal{H}'_{n_{s}}$
is the quantized field energy for the harmonic fields of equation
(\ref{eq:stationary-state-energy-bounded}) for the bounded space,
or equation (\ref{eq:stationary-state-energy-unbounded}) for the
free space. For the bounded space, the quantized field energy $\mathcal{H}'_{n_{s}}\equiv\mathcal{H}'_{n_{s},b}$
of equation (\ref{eq:stationary-state-energy-bounded}) is a function
of the wave number $k'_{i}\left(n_{i}\right),$ which in turn is a
function of the wave mode value $n_{i}$ and the boundary functional
$f_{i}\left(L_{i}\right),$ where $L_{i}$ is the gap distance in
the direction of \begin{align*}
\vec{L}_{i} & =\left[\vec{R'}_{2}\cdot\hat{e}_{i}-\vec{R'}_{1}\cdot\hat{e}_{i}\right]\hat{e}_{i}.\end{align*}
 Here $\vec{R'}_{1}$ and $\vec{R'}_{2}$ are the position vectors
for the involved boundaries. As an illustration with the two plate
configuration shown in Figure \ref{cap:casimir-plates-outlook}, $\vec{R'}_{1}$
may represent the plate positioned at $z=0$ and $\vec{R'}_{2}$ may
correspond to the plate at the position $z=d.$ When the position
of these boundaries are changing in time, the quantized field energy
$\mathcal{H}'_{n_{s}}\equiv\mathcal{H}'_{n_{s},b}$ will be modified
accordingly because the wave number functional $k'_{i}\left(n_{i}\right)$
is varying in time, \begin{align*}
\frac{dk'_{i}}{dt} & =\frac{\partial k'_{i}}{\partial n_{i}}\frac{dn_{i}}{dt}f_{i}\left(L_{i}\right)+n_{i}\frac{\partial f_{i}}{\partial L_{i}}\frac{dL_{i}}{dt}\\
 & =f_{i}\left(L_{i}\right)\frac{\partial k'_{i}}{\partial n_{i}}\dot{n}_{i}+n_{i}\frac{\partial f_{i}}{\partial L_{i}}\dot{L}_{i}.\end{align*}
 Here the term proportional to $\dot{n}_{i}$ refers to the case where
the boundaries remain fixed throughout all times but the number of
wave modes in the resonator are being driven by some active external
influence. The term proportional to $\dot{L}_{i}$ represents the
changes in the number of wave modes due to the moving boundaries. 

For an isolated system, there are no external influences, hence $\dot{n}_{i}=0.$
The dynamical force arising from the time variation of the boundaries
is found to be (see Appendix D1 of reference \cite{key-Sung-Nae-Cho}),
\begin{align}
\vec{\mathcal{F}'}_{\alpha} & =\sum_{i=1}^{3}\left\{ n_{i}\frac{\partial f_{i}}{\partial L_{i}}\left[C_{\alpha,5}\frac{\partial^{2}\mathcal{H}'_{n_{s}}}{\partial\left[k'_{i}\right]^{2}}+\left(1-\delta_{i\alpha}\right)\right.\right.\nonumber \\
 & \left.\times\left(C_{\alpha,6}-C_{\alpha,7}\left[n_{s}+\frac{1}{2}\right]k'_{i}\right)\left[n_{s}+\frac{1}{2}\right]\right]\dot{L}_{i}\nonumber \\
 & \left.+\sum_{j=1}^{3}\left(1-\delta_{ij}\right)C_{\alpha,5}n_{j}\frac{\partial f_{j}}{\partial L_{j}}\frac{\partial^{2}\mathcal{H}'_{n_{s}}}{\partial k'_{j}\partial k'_{i}}\dot{L}_{j}\right\} \hat{e_{\alpha}},\label{eq:dynamical-force-L-dot-ONLY-3D-NA}\end{align}
 where $C_{\alpha,1},$ $C_{\alpha,2},$ $C_{\alpha,3},$ $C_{\alpha,4},$
$C_{\alpha,5},$ $C_{\alpha,6}$ and $C_{\alpha,7}$ are defined as
\begin{equation}
\left\{ \begin{array}{c}
C_{\alpha,1}=\sum_{i=1}^{3}\frac{\partial\mathcal{H}'_{n_{s}}}{\partial k'_{i}},\\
\\C_{\alpha,2}=\sum_{i=1}^{3}\left(1-\delta_{i\alpha}\right)\left(\left[n_{s}+\frac{1}{2}\right]\hbar c\right)^{2}k'_{i},\\
\\C_{\alpha,3}=\sum_{i=1}^{3}\left(1-\delta_{i\alpha}\right)\left(\left[n_{s}+\frac{1}{2}\right]\hbar\right)^{2}\left[k'_{i}\right]^{2},\end{array}\right.\label{eq:dynamical-force-C1-C2-C3-DEF}\end{equation}
 \begin{align}
C_{\alpha,4} & =\left(\frac{\left[n_{s}+\frac{1}{2}\right]^{2}\hbar^{2}C_{\alpha,2}^{2}}{\left[C_{\alpha,1}^{2}-\left(\left[n_{s}+\frac{1}{2}\right]\hbar c\right)^{2}\right]^{2}}\right.\nonumber \\
 & \left.+\frac{C_{\alpha,2}^{2}-C_{\alpha,1}^{2}C_{\alpha,3}}{C_{\alpha,1}^{2}-\left(\left[n_{s}+\frac{1}{2}\right]\hbar c\right)^{2}}\right)^{-1/2},\label{eq:dynamical-force-C4-DEF}\end{align}
 \begin{align}
C_{\alpha,5} & =\frac{C_{\alpha,1}C_{\alpha,4}\left[C_{\alpha,1}^{2}C_{\alpha,3}-C_{\alpha,2}^{2}\right]}{\left[C_{\alpha,1}^{2}-\left(\left[n_{s}+\frac{1}{2}\right]\hbar c\right)^{2}\right]^{2}}\nonumber \\
 & -\frac{2\left[n_{s}+\frac{1}{2}\right]^{2}\hbar^{2}C_{\alpha,1}C_{\alpha,2}^{2}C_{\alpha,4}}{\left[C_{\alpha,1}^{2}-\left(\left[n_{s}+\frac{1}{2}\right]\hbar c\right)^{2}\right]^{3}}\nonumber \\
 & -\frac{2\left[n_{s}+\frac{1}{2}\right]\hbar C_{\alpha,1}C_{\alpha,2}}{\left[C_{\alpha,1}^{2}-\left(\left[n_{s}+\frac{1}{2}\right]\hbar c\right)^{2}\right]^{2}}\nonumber \\
 & -\frac{C_{\alpha,1}C_{\alpha,3}C_{\alpha,4}}{C_{\alpha,1}^{2}-\left(\left[n_{s}+\frac{1}{2}\right]\hbar c\right)^{2}},\label{eq:dynamical-force-C5-DEF}\end{align}
 \begin{align}
C_{\alpha,6} & =\frac{\left[n_{s}+\frac{1}{2}\right]^{2}\hbar^{2}C_{\alpha,2}C_{\alpha,4}}{\left[C_{\alpha,1}^{2}-\left(\left[n_{s}+\frac{1}{2}\right]\hbar c\right)^{2}\right]^{2}}\nonumber \\
 & +\frac{C_{\alpha,2}C_{\alpha,4}}{C_{\alpha,1}^{2}-\left(\left[n_{s}+\frac{1}{2}\right]\hbar c\right)^{2}}\nonumber \\
 & +\frac{\left[n_{s}+\frac{1}{2}\right]\hbar}{C_{\alpha,1}^{2}-\left(\left[n_{s}+\frac{1}{2}\right]\hbar c\right)^{2}},\label{eq:dynamical-force-C6-DEF}\end{align}
 \begin{align}
C_{\alpha,7} & =\frac{C_{\alpha,1}^{2}C_{\alpha,4}}{C_{\alpha,1}^{2}-\left(\left[n_{s}+\frac{1}{2}\right]\hbar c\right)^{2}}.\label{eq:dynamical-force-C7-DEF}\end{align}
 The force shown in the above expression vanishes for the one dimensional
case. This is an expected result. To understand why the force vanishes,
we have to refer to the starting point equation, \begin{align}
\sum_{i=1}^{3}\left[\left(\left[n_{s}+\frac{1}{2}\right]\hbar\right)^{-1}\frac{\partial\mathcal{H}'_{n_{s}}}{\partial k'_{i}}\right.\nonumber \\
\left.-\left\{ \sum_{i=1}^{3}\left[p'_{i}\right]^{2}\right\} ^{-1/2}cp'_{i}\right]dp'_{i} & =0,\label{eq:d-Hamilton-minus-d-momentum-mag-combined}\end{align}
 of the Appendix D1 of reference \cite{key-Sung-Nae-Cho}. The summation
here obviously runs only once to arrive at the expression, $\partial\mathcal{H}'_{n_{s}}/\partial k'_{i}=\left[n_{s}+\frac{1}{2}\right]\hbar c.$
This is a classic situation where the problem has been over specified.
For the \textbf{3D} case, equation (\ref{eq:d-Hamilton-minus-d-momentum-mag-combined})
is a combination of two constraints, $\sum_{i=1}^{3}\left[p'_{i}\right]^{2}$
and $\mathcal{H}'_{n_{s}}.$ For the one dimensional case, there is
only one constraint, $\mathcal{H}'_{n_{s}}.$ Therefore, equation
(\ref{eq:d-Hamilton-minus-d-momentum-mag-combined}) becomes an over
specification. In order to avoid the problem caused by over specifications
in this formulation, the one dimensional force expression can be obtained
directly by differentiating the equation (\ref{eq:2-3-1}) instead
of using the above formulation for the three dimensional case. The
\textbf{1D} dynamical force expression for an isolated, non-driven
systems then becomes \begin{align}
\vec{\mathcal{F}'} & =\frac{n}{c}\frac{\partial f}{\partial L}\frac{\partial\mathcal{H}'_{n_{s}}}{\partial k'}\dot{L}\hat{e},\label{eq:dynamical-force-L-dot-ONLY-1D-NA}\end{align}
 where $\vec{\mathcal{F}'}$ is an one dimensional force. Here the
subscript $\alpha$ of $\vec{\mathcal{F}'}_{\alpha}$ have been dropped
for simplicity, since it is a one dimensional force. The details of
this section can be found in Appendix D1 of reference \cite{key-Sung-Nae-Cho}.

\subsubsection{Equations of Motion for the Driven Parallel Plates}

\begin{figure}
\begin{center}\includegraphics[%
  scale=0.5]{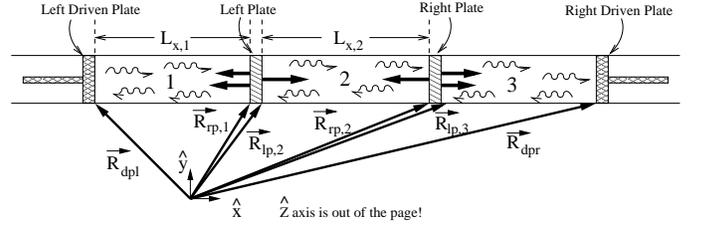}\end{center}

\caption{A one dimensional driven parallel plates configuration. \label{cap:driven-parallel-plates-force}}
\end{figure}

The Unruh-Davies effect states that heating up of an accelerating
conductor plate is proportional to its acceleration through the relation
\begin{align*}
T & =\frac{\hbar\ddot{R}}{2\pi ck'},\end{align*}
 where $\ddot{R}$ is the plate acceleration. A one dimensional system
of two parallel plates, shown in Figure \ref{cap:driven-parallel-plates-force},
can be used as a simple model to demonstrate the complicated sonoluminescense
phenomenon for a bubble subject to a strong acoustic field. 

The dynamical force for the \textbf{1D}, linear coupled system can
be expressed with equation (\ref{eq:dynamical-force-L-dot-ONLY-1D-NA}),
\begin{align}
 & \left\{ \begin{array}{c}
\ddot{R}_{1}-\eta_{1}\dot{R}_{1}-\eta_{2}\dot{R}_{2}=\xi_{rp},\\
\\\ddot{R}_{2}-\eta_{3}\dot{R}_{2}-\eta_{4}\dot{R}_{1}=\xi_{lp},\end{array}\right.\label{eq:Eq-of-Mot-Plate-System-Static-NA}\end{align}
 where \begin{equation}
\left\{ \begin{array}{c}
\eta_{1}=m_{rp}^{-1}\left[s_{rp,2}g_{\alpha,2}-s_{rp,3}g_{\alpha,3}\right],\\
\\\eta_{2}=-s_{rp,2}g_{\alpha,2}m_{rp}^{-1}\\
\\\eta_{3}=m_{lp}^{-1}\left[s_{lp,1}g_{\alpha,1}-s_{lp,2}g_{\alpha,2}\right],\\
\\\eta_{4}=s_{lp,2}g_{\alpha,2}m_{lp}^{-1},\\
\\\xi_{rp}=s_{rp,3}g_{\alpha,3}m_{rp}^{-1}\dot{R}_{dpr,cm,\alpha},\\
\\\xi_{lp}=-s_{lp,1}g_{\alpha,1}m_{lp}^{-1},\\
\\R_{1}=R_{rp,cm,\alpha},\\
\\R_{2}=R_{lp,cm,\alpha},\end{array}\right.\label{eq:eta-1-2-3-4-ci-lp-rp-DEF}\end{equation}
 \begin{eqnarray*}
\dot{R}_{a,b}=\frac{d\vec{R}_{a}}{dt}\cdot\hat{e_{b}}, &  & \ddot{R}_{a,b}=\frac{d^{2}\vec{R}_{a}}{dt^{2}}\cdot\hat{e_{b}},\end{eqnarray*}
 \begin{align}
g_{\alpha,\Re} & =\frac{n_{\alpha,\Re}}{c}\left(\frac{\partial f_{\alpha,\Re}}{\partial L_{\alpha,\Re}}\right)\left(\frac{\partial\mathcal{H}'_{n_{s},\Re}}{\partial k'_{\alpha,\Re}}\right).\label{eq:g-i-R-DEF}\end{align}
 Here $R_{1}$ represents the center of mass position for the {}``Right
Plate'' and $R_{2}$ represents the center of mass position for the
{}``Left Plate'' as illustrated in Figure \ref{cap:driven-parallel-plates-force}.
$m_{rp}$ and $m_{lp}$ are the corresponding masses for the right
and left plates. $s_{rp,i}$ and $s_{lp,i}$ are the appropriate $\left(\pm\right)$signs
that needs to be determined. For the mean time, the exact signs for
$s_{rp,i}$ and $s_{lp,i}$ are not relevant, hence it is left as
is. With a slight modification, equation (\ref{eq:Eq-of-Mot-Plate-System-Static-NA})
for this linear coupled system can be written in the matrix form,
(see Appendix D2 of reference \cite{key-Sung-Nae-Cho}): \begin{eqnarray*}
R_{1}=\int_{t_{0}}^{t}R_{3}dt', &  & R_{2}=\int_{t_{0}}^{t}R_{4}dt',\end{eqnarray*}
 and \begin{align}
\underbrace{\left[\begin{array}{c}
\dot{R}_{3}\\
\dot{R}_{4}\end{array}\right]}_{\dot{\vec{R}}_{\eta}} & =\underbrace{\left[\begin{array}{cc}
\eta_{1} & \eta_{2}\\
\eta_{4} & \eta_{3}\end{array}\right]}_{\widetilde{M}_{\eta}}\underbrace{\left[\begin{array}{c}
R_{3}\\
R_{4}\end{array}\right]}_{\vec{R}_{\eta}}+\underbrace{\left[\begin{array}{c}
\xi_{rp}\\
\xi_{lp}\end{array}\right]}_{\vec{\xi}},\label{eq:Eq-of-Mot-Plate-System-Static-Matrix-NA}\end{align}
 where \[
\left\{ \begin{array}{c}
\begin{array}{ccc}
\dot{R}_{1}=R_{3}, &  & \dot{R}_{2}=R_{4},\\
\end{array}\\
\dot{R}_{3}=\ddot{R}_{1}=\xi_{rp}+\eta_{1}\dot{R}_{1}+\eta_{2}\dot{R}_{2}=\xi_{rp}+\eta_{1}R_{3}+\eta_{2}R_{4},\\
\\\dot{R}_{4}=\ddot{R}_{2}=\xi_{lp}+\eta_{3}\dot{R}_{2}+\eta_{4}\dot{R}_{1}=\xi_{lp}+\eta_{3}R_{4}+\eta_{4}R_{3}.\end{array}\right.\]
The matrix equation has the solutions (see Appendix D2 of reference
\cite{key-Sung-Nae-Cho}): \begin{align}
 & \dot{R}_{rp,cm,\alpha}\left(t\right)=\left[\frac{\lambda_{4}\left(;t_{0}\right)-\eta_{1}\left(;t_{0}\right)}{\lambda_{3}\left(;t_{0}\right)-\eta_{1}\left(;t_{0}\right)}-1\right]^{-1}\nonumber \\
 & \times\frac{\psi_{11}\left(t,t_{0}\right)\dot{R}_{rp,cm,\alpha}\left(t_{0}\right)+\psi_{12}\left(t,t_{0}\right)\dot{R}_{lp,cm,\alpha}\left(t_{0}\right)}{\exp\left(\left[\lambda_{3}\left(;t_{0}\right)+\lambda_{4}\left(;t_{0}\right)\right]t_{0}\right)}\nonumber \\
 & +\int_{t_{0}}^{t}\frac{\psi_{22}\left(t',t_{0}\right)\xi_{rp}\left(t'\right)-\psi_{12}\left(t',t_{0}\right)\xi_{lp}\left(t'\right)}{\psi_{11}\left(t',t_{0}\right)\psi_{22}\left(t',t_{0}\right)-\psi_{12}\left(t',t_{0}\right)\psi_{21}\left(t',t_{0}\right)}dt'\nonumber \\
 & \times\psi_{11}\left(t,t_{0}\right)+\psi_{12}\left(t,t_{0}\right)\nonumber \\
 & \times\int_{t_{0}}^{t}\frac{\psi_{11}\left(t',t_{0}\right)\xi_{lp}\left(t'\right)-\psi_{21}\left(t',t_{0}\right)\xi_{rp}\left(t'\right)}{\psi_{11}\left(t',t_{0}\right)\psi_{22}\left(t',t_{0}\right)-\psi_{12}\left(t',t_{0}\right)\psi_{21}\left(t',t_{0}\right)}dt',\label{eq:Eq-of-Mot-R1-Speed-NA}\end{align}
 \begin{align}
 & \dot{R}_{lp,cm,\alpha}\left(t\right)=\left[\frac{\lambda_{4}\left(;t_{0}\right)-\eta_{1}\left(;t_{0}\right)}{\lambda_{3}\left(;t_{0}\right)-\eta_{1}\left(;t_{0}\right)}-1\right]^{-1}\nonumber \\
 & \times\frac{\psi_{21}\left(t,t_{0}\right)\dot{R}_{rp,cm,\alpha}+\psi_{22}\left(t,t_{0}\right)\dot{R}_{lp,cm,\alpha}\left(t_{0}\right)}{\exp\left(\left[\lambda_{3}\left(;t_{0}\right)+\lambda_{4}\left(;t_{0}\right)\right]t_{0}\right)}\nonumber \\
 & +\int_{t_{0}}^{t}\frac{\psi_{22}\left(t',t_{0}\right)\xi_{rp}\left(t'\right)-\psi_{12}\left(t',t_{0}\right)\xi_{lp}\left(t'\right)}{\psi_{11}\left(t',t_{0}\right)\psi_{22}\left(t',t_{0}\right)-\psi_{12}\left(t',t_{0}\right)\psi_{21}\left(t',t_{0}\right)}dt'\nonumber \\
 & \times\psi_{21}\left(t,t_{0}\right)+\psi_{22}\left(t,t_{0}\right)\nonumber \\
 & \times\int_{t_{0}}^{t}\frac{\psi_{11}\left(t',t_{0}\right)\xi_{lp}\left(t'\right)-\psi_{21}\left(t',t_{0}\right)\xi_{rp}\left(t'\right)}{\psi_{11}\left(t',t_{0}\right)\psi_{22}\left(t',t_{0}\right)-\psi_{12}\left(t',t_{0}\right)\psi_{21}\left(t',t_{0}\right)}dt',\label{eq:Eq-of-Mot-R2-Speed-NA}\end{align}
 where \begin{align}
\lambda_{3} & =\frac{\eta_{1}+\eta_{3}}{2}+\sqrt{\frac{1}{4}\left[\eta_{1}-\eta_{3}\right]^{2}+\eta_{2}\eta_{4}},\label{eq:Lambda-3}\end{align}
 \begin{align}
\lambda_{4} & =\frac{\eta_{1}+\eta_{3}}{2}-\sqrt{\frac{1}{4}\left[\eta_{1}-\eta_{3}\right]^{2}+\eta_{2}\eta_{4}},\label{eq:Lambda-4}\end{align}
 \begin{align}
\psi_{11}\left(t,t_{0}\right) & =\grave{R}_{3}\grave{R}_{4}\left[\frac{\lambda_{4}-\eta_{1}}{\lambda_{3}-\eta_{1}}\exp\left(\lambda_{3}t+\lambda_{4}t_{0}\right)\right.\nonumber \\
 & \left.-\exp\left(\lambda_{4}t+\lambda_{3}t_{0}\right)\right],\label{eq:Principal-Matrix-Psi-11}\end{align}
 \begin{align}
\psi_{12}\left(t,t_{0}\right) & =\grave{R}_{3}\grave{R}_{4}\left[\frac{\eta_{2}}{\lambda_{3}-\eta_{1}}\exp\left(\lambda_{4}t+\lambda_{3}t_{0}\right)\right.\nonumber \\
 & \left.-\frac{\eta_{2}}{\lambda_{3}-\eta_{1}}\exp\left(\lambda_{3}t+\lambda_{4}t_{0}\right)\right],\label{eq:Principal-Matrix-Psi-12}\end{align}
 \begin{align}
\psi_{21}\left(t,t_{0}\right) & =\grave{R}_{3}\grave{R}_{4}\left[\frac{\lambda_{4}-\eta_{1}}{\eta_{2}}\exp\left(\lambda_{3}t+\lambda_{4}t_{0}\right)\right.\nonumber \\
 & \left.-\frac{\lambda_{4}-\eta_{1}}{\eta_{2}}\exp\left(\lambda_{4}t+\lambda_{3}t_{0}\right)\right],\label{eq:Principal-Matrix-Psi-21}\end{align}
 \begin{align}
\psi_{22}\left(t,t_{0}\right) & =\grave{R}_{3}\grave{R}_{4}\left[\frac{\lambda_{4}-\eta_{1}}{\lambda_{3}-\eta_{1}}\exp\left(\lambda_{4}t+\lambda_{3}t_{0}\right)\right.\nonumber \\
 & \left.-\exp\left(\lambda_{3}t+\lambda_{4}t_{0}\right)\right].\label{eq:Principal-Matrix-Psi-22}\end{align}
 The quantities $\dot{R}_{rp,cm,\alpha}$ and $\dot{R}_{lp,cm,\alpha}$
are the speed of the center of mass of {}``Right Plate'' and the
speed of the center of mass of the {}``Left Plate,'' respectively,
and $\alpha$ defines the particular basis direction. The corresponding
positions $R_{rp,cm,\alpha}\left(t\right)$ and $R_{lp,cm,\alpha}\left(t\right)$
are found by integrating equations (\ref{eq:Eq-of-Mot-R1-Speed-NA})
and (\ref{eq:Eq-of-Mot-R2-Speed-NA}) with respect to time, \begin{align}
 & R_{rp,cm,\alpha}\left(t\right)=\left[\frac{\lambda_{4}\left(;t_{0}\right)-\eta_{1}\left(;t_{0}\right)}{\lambda_{3}\left(;t_{0}\right)-\eta_{1}\left(;t_{0}\right)}-1\right]^{-1}\nonumber \\
 & \times\int_{t_{0}}^{t}\left[\frac{\psi_{11}\left(\tau,t_{0}\right)\dot{R}_{rp,cm,\alpha}\left(t_{0}\right)+\psi_{12}\left(\tau,t_{0}\right)\dot{R}_{lp,cm,\alpha}\left(t_{0}\right)}{\exp\left(\left[\lambda_{3}\left(;t_{0}\right)+\lambda_{4}\left(;t_{0}\right)\right]t_{0}\right)}\right.\nonumber \\
 & +\int_{t_{0}}^{\tau}\frac{\psi_{22}\left(t',t_{0}\right)\xi_{rp}\left(t'\right)-\psi_{12}\left(t',t_{0}\right)\xi_{lp}\left(t'\right)}{\psi_{11}\left(t',t_{0}\right)\psi_{22}\left(t',t_{0}\right)-\psi_{12}\left(t',t_{0}\right)\psi_{21}\left(t',t_{0}\right)}dt'\nonumber \\
 & \times\psi_{11}\left(\tau,t_{0}\right)+\psi_{12}\left(\tau,t_{0}\right)\nonumber \\
 & \left.\times\int_{t_{0}}^{\tau}\frac{\psi_{11}\left(t',t_{0}\right)\xi_{lp}\left(t'\right)-\psi_{21}\left(t',t_{0}\right)\xi_{rp}\left(t'\right)}{\psi_{11}\left(t',t_{0}\right)\psi_{22}\left(t',t_{0}\right)-\psi_{12}\left(t',t_{0}\right)\psi_{21}\left(t',t_{0}\right)}dt'\right]\nonumber \\
 & \times d\tau+R_{rp,cm,\alpha}\left(t_{0}\right),\label{eq:Eq-of-Mot-R1-Position-NA}\end{align}
 \begin{align}
 & R_{lp,cm,\alpha}\left(t\right)=\left[\frac{\lambda_{4}\left(;t_{0}\right)-\eta_{1}\left(;t_{0}\right)}{\lambda_{3}\left(;t_{0}\right)-\eta_{1}\left(;t_{0}\right)}-1\right]^{-1}\nonumber \\
 & \times\int_{t_{0}}^{t}\left[\frac{\psi_{21}\left(\tau,t_{0}\right)\dot{R}_{rp,cm,\alpha}\left(t_{0}\right)+\psi_{22}\left(\tau,t_{0}\right)\dot{R}_{lp,cm,\alpha}\left(t_{0}\right)}{\exp\left(\left[\lambda_{3}\left(;t_{0}\right)+\lambda_{4}\left(;t_{0}\right)\right]t_{0}\right)}\right.\nonumber \\
 & +\int_{t_{0}}^{\tau}\frac{\psi_{22}\left(t',t_{0}\right)\xi_{rp}\left(t'\right)-\psi_{12}\left(t',t_{0}\right)\xi_{lp}\left(t'\right)}{\psi_{11}\left(t',t_{0}\right)\psi_{22}\left(t',t_{0}\right)-\psi_{12}\left(t',t_{0}\right)\psi_{21}\left(t',t_{0}\right)}dt'\nonumber \\
 & \times\psi_{21}\left(\tau,t_{0}\right)+\psi_{22}\left(\tau,t_{0}\right)\nonumber \\
 & \left.\times\int_{t_{0}}^{\tau}\frac{\psi_{11}\left(t',t_{0}\right)\xi_{lp}\left(t'\right)-\psi_{21}\left(t',t_{0}\right)\xi_{rp}\left(t'\right)}{\psi_{11}\left(t',t_{0}\right)\psi_{22}\left(t',t_{0}\right)-\psi_{12}\left(t',t_{0}\right)\psi_{21}\left(t',t_{0}\right)}dt'\right]\nonumber \\
 & \times d\tau+R_{lp,cm,\alpha}\left(t_{0}\right).\label{eq:Eq-of-Mot-R2-Position-NA}\end{align}
 The remaining integrations are straightforward and the explicit forms
will not be shown here. 

One may argue that for the static case, $\dot{R}_{rp,cm,\alpha}\left(t_{0}\right)$
and $\dot{R}_{lp,cm,\alpha}\left(t_{0}\right)$ must be zero because
the conductors seem to be fixed in position. This argument is flawed,
for any wall totally fixed in position upon impact would require an
infinite amount of energy. One has to consider the conservation of
momentum simultaneously. The wall has to have moved by the amount
$\bigtriangleup R_{wall}=\dot{R}_{wall}\triangle t,$ where $\triangle t$
is the total duration of impact, and $\dot{R}_{wall}$ is calculated
from the momentum conservation and it is non-zero. The same argument
can be applied to the apparatus shown in Figure \ref{cap:driven-parallel-plates-force}.
For that system \begin{align*}
\left\Vert \vec{p}_{virtual-photon}\right\Vert  & =\frac{1}{c}\mathcal{H}'_{n_{s},\Re}\left(t_{0}\right)\end{align*}
 and \[
\left\{ \begin{array}{c}
\dot{R}_{rp,cm,\alpha}\left(t_{0}\right)=\left\Vert \dot{\vec{R}}_{lp,3}\left(t_{0}\right)+\dot{\vec{R}}_{rp,2}\left(t_{0}\right)\right\Vert ,\\
\\\dot{R}_{lp,cm,\alpha}\left(t_{0}\right)=\left\Vert \dot{\vec{R}}_{rp,1}\left(t_{0}\right)+\dot{\vec{R}}_{lp,2}\left(t_{0}\right)\right\Vert .\end{array}\right.\]
 For simplicity, assuming that the impact is always only in the normal
direction, \begin{align*}
\dot{R}_{rp,cm,\alpha}\left(t_{0}\right) & =\frac{2}{m_{rp}c}\left\Vert \mathcal{H}'_{n_{s},3}\left(t_{0}\right)-\mathcal{H}'_{n_{s},2}\left(t_{0}\right)\right\Vert \end{align*}
 and \begin{align*}
\dot{R}_{lp,cm,\alpha}\left(t_{0}\right) & =\frac{2}{m_{lp}c}\left\Vert \mathcal{H}'_{n_{s},1}\left(t_{0}\right)-\mathcal{H}'_{n_{s},2}\left(t_{0}\right)\right\Vert ,\end{align*}
 where the differences under the magnitude symbol imply field energies
from different regions counteract the other. The details of this section
can be found in Appendix D2 of reference \cite{key-Sung-Nae-Cho}.

\section{Results and Outlook}

\begin{figure}
\begin{center}\includegraphics[%
  scale=0.5]{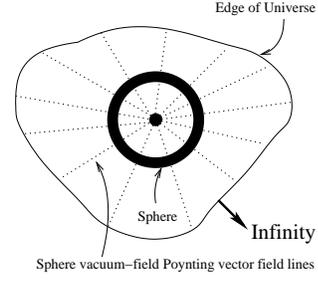}\end{center}

\caption{Boyer's configuration is such that a sphere is the only matter in
the entire universe. His universe extends to the infinity, hence there
are no boundaries. The sense of vacuum-field energy flow is along
the radial vector $\hat{r},$ which is defined with respect to the
sphere center. \label{cap:boyers-sphere-in-universe}}
\end{figure}

The results for the sign of Casimir force on non-planar geometric
configurations considered in this investigation will eventually be
compared with the classic repulsive result obtained by Boyer decades
earlier. For this reason, it is worth reviewing Boyer's original configuration
as shown in Figure \ref{cap:boyers-sphere-in-universe}.

T. H. Boyer in 1968 obtained a repulsive Casimir force result for
his charge-neutral, hollow spherical shell of a perfect conductor
\cite{key-Boyer}. For simplicity, his sphere is the only object in
the entire universe and, therefore, no external boundaries such as
laboratory walls, etc., were defined in his problem. Furthermore,
the zero-point energy flow is always perpendicular to his sphere.
Such restriction constitutes a very stringent condition for the material
property that a sphere has to meet. For example, if one were to look
at Boyer's sphere, he would not see the whole sphere; but instead,
he would see a small spot on the surface of a sphere that happens
to be in his line of sight. This happens because the sphere in Boyer's
configuration can only radiate in a direction normal to the surface.
One could equivalently argue that Boyer's sphere only responds to
the approaching radiation at normal angles of incidence with respect
to the surface of the sphere. When the Casimir force is computed for
such restricted radiation energy flow, the result is repulsive. In
Boyer's picture, this may be attributed to the fact that closer to
the origin of a sphere, the spherically symmetric radiation energy
flow becomes more dense due to the inverse length dependence, and
this density decreases as it gets further away from the sphere center.
This argument, however, seems to be flawed because it inherently implies
existence of the preferred origin for the vacuum fields. As an illustration,
Boyer's sphere is shown in Figure \ref{cap:boyers-sphere-in-universe}.
For the rest of this investigation, {}``Boyer's sphere'' would be
strictly referred to as the sphere made of material with such a property
that it only radiates or responds to vacuum-field radiations at normal
angle of incidence with respect to its surface.

\begin{figure}
\begin{center}\includegraphics[%
  scale=0.4]{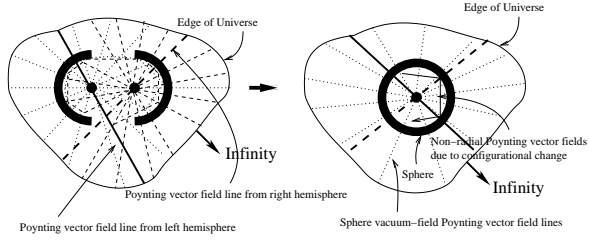}\end{center}

\caption{Manufactured sphere, in which two hemispheres are brought together,
results in small non-spherically symmetric vacuum-field radiation
inside the cavity due to the configuration change. For the hemispheres
made of Boyer's material, these fields in the resonator will eventually
get absorbed by the conductor resulting in heating of the hemispheres.
\label{cap:boyers-sphere-in-universe-hemi-to-sphere}}
\end{figure}

The formation of a sphere by bringing together two nearby hemispheres
satisfying the material property of Boyer's sphere is illustrated
in Figure \ref{cap:boyers-sphere-in-universe-hemi-to-sphere}. Since
Boyer's material property only allow radiation in the normal direction
to its surface, the radiation associated with each hemisphere would
necessarily go through the corresponding hemisphere centers. For clarity,
let us define the unit radial basis vector associated with the left
and right hemispheres by $\hat{r}_{L}$ and $\hat{r}_{R},$ respectively.
If the hemispheres are made of normal conductors the radiation from
one hemisphere entering the other hemisphere cavity would go through
a complex series of reflections before escaping the cavity. Here,
a conductor with Boyer's stringent material property is not considered
normal. Conductors that are normal also radiate in directions non-normal
to their surface, whereas Boyer's conductor can only radiate normal
to its surface. Due to the fact that Boyer's conducting materials
can only respond to radiation impinging at a normal angle of incidence
with respect to its surface, all of the incoming radiation at oblique
angles of incidence with respect to the local surface normal is absorbed
by the host hemisphere. This suggests that for the hemisphere-hemisphere
arrangement made of Boyer's material shown in Figure \ref{cap:boyers-sphere-in-universe-hemi-to-sphere},
the temperature of the two hemispheres would rise indefinitely over
time. This does not happen with ordinary conductors. This suggests
that Boyer's conducting material, of which his sphere is made, is
completely hypothetical. Precisely because of this material assumption,
Boyer's Casimir force is repulsive. 

\begin{figure}
\begin{center}\includegraphics[%
  scale=0.5]{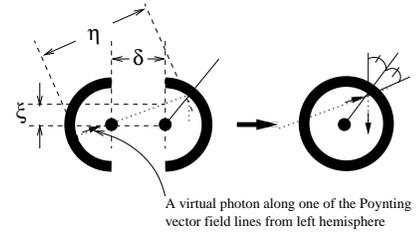}\end{center}

\caption{The process in which a configuration change from hemisphere-hemisphere
to sphere inducing virtual photon in the direction other than $\hat{r}$
is shown. The virtual photon here is referred to as the quanta of
energy associated with the zero-point radiation. \label{cap:boyers-sphere-in-universe-Config-Demo}}
\end{figure}

For the moment, let us relax the stringent Boyer's material property
for the hemispheres to that of ordinary conductors. For the hemispheres
made of ordinary conducting materials, there would result a series
of reflections in one hemisphere cavity due to those radiations entering
the cavity from nearby hemisphere. For simplicity, the ordinary conducting
material referred to here is that of perfect conductors without Boyer's
hypothetical material property requirement. Furthermore, only the
radiation emanating normally with respect to its surface is considered.
The idea is to illustrate that the {}``normally emanated radiation''
from one hemisphere results in elaboration of the effects of {}``obliquely
emanated radiation'' on another hemisphere cavity. Here the obliquely
emanated radiation means those radiation emanating from a surface
not along the local normal of the surface. 

When two such hemispheres are brought together to form a sphere, there
would exist some radiation trapped in the sphere of which the radiation
energy flow lines are not spherically symmetric with respect to the
sphere center. To see how a mere change in configuration invokes such
non-spherically symmetric energy flow, consider the illustration shown
in Figure \ref{cap:boyers-sphere-in-universe-Config-Demo}. For clarity,
only one {}``normally emanated radiation'' energy flow line from
the left hemisphere is shown. When one brings together the two hemispheres
just in time before that quantum of energy escapes the hemisphere
cavity to the right, the trapped energy quantum would continuously
go through series of complex reflections in the cavity obeying the
reflection law. But how fast or how slow one brings in two hemispheres
is irrelevant in invoking such non-spherically symmetric energy flow
because the gap $\delta$ can be chosen arbitrarily. Therefore, there
would always be a stream of energy quanta crossing the hemisphere
opening with $\xi\neq0$ as shown in Figure \ref{cap:boyers-sphere-in-universe-Config-Demo}.
In other words, there is always a time interval $\triangle t$ within
which the hemispheres are separated by an amount $\delta$ before
closure. The quanta of vacuum-field radiation energy created within
that time interval $\triangle t$ would always be satisfying the condition
$\xi\neq0,$ and this results in reflections at oblique angle of incidence
with respect to the local normal of the walls of inner sphere cavity.
Only when the two hemispheres are finally closed, would then $\xi=0$
and the radiation energy produced in the sphere after that moment
would be spherically symmetric and the reflections would be normal
to the surface. However, those trapped quantum of energy that were
produced prior to the closure of the two hemispheres would always
be reflecting from the inner sphere surface at oblique angles of incidence. 

\begin{figure}
\begin{center}\includegraphics[%
  scale=0.5]{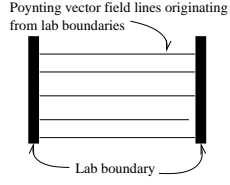}\end{center}

\caption{A realistic laboratory has boundaries, e.g., walls. These boundaries
have effect similar to the field modes between two parallel plates.
In \textbf{3D}, the effects are similar to that of a cubical laboratory,
etc.  \label{cap:reallab}}
\end{figure}

Unlike Boyer's ideal laboratory, realistic laboratories have boundaries
made of ordinary material as illustrated in Figure \ref{cap:reallab}.
One must then take into account, when calculating the Casimir force,
the vacuum-field radiation pressure contributions from the involved
conductors, as well as those contributions from the boundaries such
as laboratory walls, etc. We will examine the physics of placing two
hemispheres inside the laboratory. 

\begin{figure}
\begin{center}\includegraphics[%
  scale=0.4]{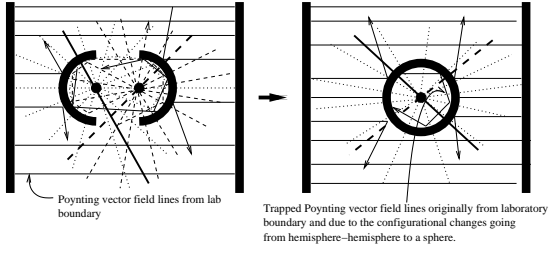}\end{center}

\caption{The schematic of sphere manufacturing process in a realistic laboratory.
\label{cap:boyers-sphere-in-reallab}}
\end{figure}

For simplicity, the boundaries of the laboratory as shown in Figure
\ref{cap:boyers-sphere-in-reallab} are assumed to be simple cubical.
Normally, the dimension of conductors considered in Casimir force
experiment is in the ranges of microns. When this is compared with
the size of the laboratory boundaries such as the walls, the walls
of the laboratory can be treated as a set of infinite parallel plates
and the vacuum-fields inside the the laboratory can be treated as
simple plane waves with impunity. 

The presence of laboratory boundaries induce reflection of energy
flow similar to that between the two parallel plate arrangement. When
the two hemisphere arrangement shown in Figure \ref{cap:boyers-sphere-in-universe-hemi-to-sphere}
is placed in such a laboratory, the result is to elaborate the radiation
pressure contributions from obliquely incident radiations on external
surfaces of the two hemispheres. If the two hemispheres are made of
conducting material satisfying Boyer's material property, the vacuum-field
radiation impinging on hemisphere surfaces at oblique angles of incidence
would cause heating of the hemispheres. It means that Boyer's hemispheres
placed in a realistic laboratory would continue to rise in temperature
as a function of time. However, this does not happen with ordinary
conductors.

If the two hemispheres are made of ordinary perfect conducting materials,
the reflections of radiation at oblique angles of incidence from the
laboratory boundaries would elaborate on the radiation pressure acting
on the external surfaces of two hemispheres at oblique angles of incidence.
Because Boyer's sphere only radiates in the normal direction to its
surface, or only responds to impinging radiation at normal incidence
with respect to the sphere surface, the extra vacuum-field radiation
pressures considered here, i.e., the ones involving oblique angles
of incidence, are missing in his Casimir force calculation for the
sphere.

\subsection{Results}

T. H. Boyer in 1968 have shown that for a charge-neutral, perfect
conductor of hollow spherical shell, the sign of the Casimir force
is positive, which means the force is repulsive. He reached this conclusion
by assuming that all vacuum-field radiation energy flows for his sphere
are spherically symmetric with respect to its center. In other words,
only the wave vectors that are perpendicular to his sphere surface
were included in the Casimir force calculation. In the following sections,
the non-perpendicular wave vector contributions to the Casimir force
that were not accounted for in Boyer's work are considered.

\subsubsection{Hollow Spherical Shell}

\begin{figure}
\begin{center}\includegraphics[%
  scale=0.5]{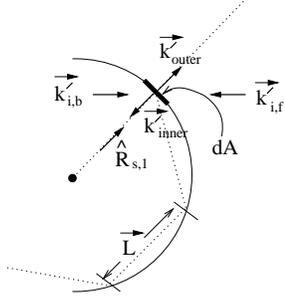}\end{center}

\caption{The vacuum-field wave vectors $\vec{k'}_{i,b}$ and $\vec{k'}_{i,f}$
impart a net momentum of the magnitude $\left\Vert \vec{p}_{net}\right\Vert =\hbar\left\Vert \vec{k'}_{i,b}-\vec{k'}_{i,f}\right\Vert /2$
on differential patch of an area $dA$ on a conducting spherical surface.
\label{cap:effective-momentum}}
\end{figure}

As shown in Figure \ref{cap:effective-momentum}, the vacuum-field
radiation imparts upon a differential patch of an area $dA$ on the
inner wall of the conducting spherical cavity a net momentum of the
amount \begin{align*}
\triangle\vec{p}_{inner} & =-\frac{1}{2}\hbar\triangle\vec{k'}_{inner}\left(;\vec{R'}_{s,1},\vec{R'}_{s,0}\right)\\
 & =\frac{2n\pi\hbar\cos\theta_{inc}}{\left\Vert \vec{R}_{s,2}\left(r'_{i},\vec{\Lambda}'_{s,2}\right)-\vec{R}_{s,1}\left(r'_{i},\vec{\Lambda}'_{s,1}\right)\right\Vert }\hat{R'}_{s,1},\end{align*}
 where \[
\left\{ \begin{array}{c}
0\leq\theta_{inc}<\pi/2,\\
\\n=1,2,3,\cdots.\end{array}\right.\]
 Here $\triangle\vec{k'}_{inner}\left(;\vec{R'}_{s,1},\vec{R'}_{s,0}\right)$
is from equation (\ref{eq:SPHERE-Delta-K-Inside-NA}). The angle of
incidence $\theta_{inc}$ is from equation (\ref{eq:angle-of-incidence-exp});
$\vec{R}_{s,1}\left(r'_{i},\vec{\Lambda}'_{s,1}\right)$ and $\vec{R}_{s,2}\left(r'_{i},\vec{\Lambda}'_{s,2}\right)$
follow the generic form shown in equation (\ref{eq:Nth-Reflection-Point-sphere-NA}). 

Similarly, the vacuum-field radiation imparts upon a differential
patch of an area $dA$ on the outer surface of the conducting spherical
shell a net momentum of the amount \begin{align*}
\triangle\vec{p}_{outer} & =-\frac{1}{2}\hbar\triangle\vec{k'}_{outer}\left(;\vec{R'}_{s,1}+a\hat{R'}_{s,1}\right)\\
 & =-2\hbar\left\Vert \vec{k'}_{i,f}\right\Vert \cos\theta_{inc}\hat{R'}_{s,1},\end{align*}
 where \[
\left\{ \begin{array}{c}
0\leq\theta_{inc}<\pi/2,\\
\\n=1,2,3,\cdots.\end{array}\right.\]
 Here $\triangle\vec{k'}_{outer}\left(;\vec{R'}_{s,1}+a\hat{R'}_{s,1}\right)$
is from equation (\ref{eq:SPHERE-Delta-K-Outside-NA}). 

The net average force per unit time, per initial wave vector direction,
acting on differential element patch of an area $dA$ is given by
\begin{align*}
\vec{\mathcal{F}}_{s,avg} & =\lim_{\triangle t\rightarrow1}\left(\frac{\triangle\vec{p}_{outer}}{\triangle t}+\frac{\triangle\vec{p}_{inner}}{\triangle t}\right)\end{align*}
 or \begin{align*}
\vec{\mathcal{F}}_{s,avg} & =\left[\frac{n\pi}{\left\Vert \vec{R}_{s,2}\left(r'_{i},\vec{\Lambda}'_{s,2}\right)-\vec{R}_{s,1}\left(r'_{i},\vec{\Lambda}'_{s,1}\right)\right\Vert }-\left\Vert \vec{k'}_{i,f}\right\Vert \right]\\
 & \times2\hbar\cos\theta_{inc}\hat{R'}_{s,1},\end{align*}
 where \[
\left\{ \begin{array}{c}
0\leq\theta_{inc}<\pi/2,\\
\\n=1,2,3,\cdots.\end{array}\right.\]
 Notice that $\vec{\mathcal{F}}_{s,avg}$ is called a force per initial
wave vector direction because it is computed for $\vec{k'}_{i,b}$
and $\vec{k'}_{i,f}$ along specific initial directions. Here $\vec{k'}_{i,b}$
denotes a particular initial wave vector $\vec{k'}_{i}$ entering
the resonator at $\vec{R}_{s,0}$ as shown in Figure \ref{cap:sphere-reflection-dynamics}.
The subscript $b$ for $\vec{k'}_{i,b}$ denotes the bounded space
inside the resonator. The $\vec{k'}_{i,f}$ denotes a particular initial
wave vector $\vec{k'}_{i}$ impinging upon the surface of the unbounded
region of sphere at point $\vec{R'}_{s,1}+a\hat{R'}_{s,1}$ as shown
in Figure \ref{cap:sphere-reflection-dynamics}. The subscript $f$
for $\vec{k'}_{i,f}$ denotes the free space external to the resonator. 

Because the wave vector $\vec{k'}_{i,f}$ resides in free (unbounded)
space, its magnitude $\left\Vert \vec{k'}_{i,f}\right\Vert $ can
take on a continuum of allowed modes. The wave vector $\vec{k'}_{i,b}$
however resides in bounded region, hence $\left\Vert \vec{k'}_{i,b}\right\Vert $
is restricted by the relation \begin{align*}
\left\Vert \vec{L}\right\Vert  & =\left\Vert \vec{R}_{s,2}\left(r'_{i},\vec{\Lambda}'_{s,2}\right)-\vec{R}_{s,1}\left(r'_{i},\vec{\Lambda}'_{s,1}\right)\right\Vert .\end{align*}
 The free space limit is the case where the radius of the sphere becomes
very large. Therefore, by designating $\left\Vert \vec{k'}_{i,f}\right\Vert $
as \begin{align*}
\left\Vert \vec{k'}_{i,f}\right\Vert  & =\lim_{r'_{i}\rightarrow\infty}\frac{n\pi}{\left\Vert \vec{R}_{s,2}\left(r'_{i},\vec{\Lambda}'_{s,2}\right)-\vec{R}_{s,1}\left(r'_{i},\vec{\Lambda}'_{s,1}\right)\right\Vert },\end{align*}
 and summing over all allowed modes, the total average force per unit
time, per initial wave vector direction, per unit area is given by
\begin{align*}
\vec{\mathcal{F}}_{s,avg} & =\hat{R'}_{s,1}\left[\sum_{n=1}^{\infty}\frac{n\pi2\hbar\cos\theta_{inc}}{\left\Vert \vec{R}_{s,2}\left(r'_{i},\vec{\Lambda}'_{s,2}\right)-\vec{R}_{s,1}\left(r'_{i},\vec{\Lambda}'_{s,1}\right)\right\Vert }\right.\\
 & \left.-\lim_{r'_{i}\rightarrow\infty}\sum_{n=1}^{\infty}\frac{n\pi2\hbar\cos\theta_{inc}}{\left\Vert \vec{R}_{s,2}\left(r'_{i},\vec{\Lambda}'_{s,2}\right)-\vec{R}_{s,1}\left(r'_{i},\vec{\Lambda}'_{s,1}\right)\right\Vert }\right].\end{align*}
 In the limit $r'_{i}\rightarrow\infty,$ the second summation to
the right can be replaced by an integration, $\sum_{n=1}^{\infty}\rightarrow\int_{0}^{\infty}dn.$
Hence, we have \begin{align*}
\vec{\mathcal{F}}_{s,avg} & =\hat{R'}_{s,1}\left[\sum_{n=1}^{\infty}\frac{2\hbar n\pi\cos\theta_{inc}}{\left\Vert \vec{R}_{s,2}\left(r'_{i},\vec{\Lambda}'_{s,2}\right)-\vec{R}_{s,1}\left(r'_{i},\vec{\Lambda}'_{s,1}\right)\right\Vert }\right.\\
 & \left.-\lim_{r'_{i}\rightarrow\infty}\int_{0}^{\infty}\frac{2\hbar n\pi\cos\theta_{inc}dn}{\left\Vert \vec{R}_{s,2}\left(r'_{i},\vec{\Lambda}'_{s,2}\right)-\vec{R}_{s,1}\left(r'_{i},\vec{\Lambda}'_{s,1}\right)\right\Vert }\right],\end{align*}
 or with the following substitutions, \begin{align*}
k'_{i,f} & \equiv\frac{n\pi}{\left\Vert \vec{R}_{s,2}\left(r'_{i},\vec{\Lambda}'_{s,2}\right)-\vec{R}_{s,1}\left(r'_{i},\vec{\Lambda}'_{s,1}\right)\right\Vert },\end{align*}
 \begin{align*}
dn & =\frac{1}{\pi}\left\Vert \vec{R}_{s,2}\left(r'_{i},\vec{\Lambda}'_{s,2}\right)-\vec{R}_{s,1}\left(r'_{i},\vec{\Lambda}'_{s,1}\right)\right\Vert dk'_{i,f},\end{align*}
 the total average force per unit time, per initial wave vector direction,
per unit area is written as \begin{align}
\vec{\mathcal{F}}_{s,avg} & =\left[\sum_{n=1}^{\infty}\frac{n\pi}{\left\Vert \vec{R}_{s,2}\left(r'_{i},\vec{\Lambda}'_{s,2}\right)-\vec{R}_{s,1}\left(r'_{i},\vec{\Lambda}'_{s,1}\right)\right\Vert }\right.\nonumber \\
 & -\frac{1}{\pi}\lim_{r'_{i}\rightarrow\infty}\left(\left\Vert \vec{R}_{s,2}\left(r'_{i},\vec{\Lambda}'_{s,2}\right)-\vec{R}_{s,1}\left(r'_{i},\vec{\Lambda}'_{s,1}\right)\right\Vert \right.\nonumber \\
 & \left.\left.\times\int_{0}^{\infty}k'_{i,f}dk'_{i,f}\right)\right]2\hbar\cos\theta_{inc}\hat{R'}_{s,1},\label{eq:sphere-ambient-force-average}\end{align}
 where $0\leq\theta_{inc}<\pi/2$ and $n=1,2,3,\cdots.$ The total
average vacuum-field radiation force per unit time acting on the uncharged
conducting spherical shell is therefore\begin{align*}
\vec{F}_{s,total} & =\sum_{\left\{ \vec{k'}_{i,b},\vec{k'}_{i,f},\vec{R'}_{s,0}\right\} }\int_{S}\vec{\mathcal{F}}_{s,avg}\cdot d\vec{S}_{sphere}\end{align*}
 or \begin{align}
\vec{F}_{s,total} & =\sum_{\left\{ \vec{k'}_{i,b},\vec{k'}_{i,f},\vec{R'}_{s,0}\right\} }\int_{S}\nonumber \\
 & \times\left[\sum_{n=1}^{\infty}\frac{2n\pi\hbar\cos\theta_{inc}}{\left\Vert \vec{R}_{s,2}\left(r'_{i},\vec{\Lambda}'_{s,2}\right)-\vec{R}_{s,1}\left(r'_{i},\vec{\Lambda}'_{s,1}\right)\right\Vert }\right.\nonumber \\
 & -\lim_{r'_{i}\rightarrow\infty}\left(\left\Vert \vec{R}_{s,2}\left(r'_{i},\vec{\Lambda}'_{s,2}\right)-\vec{R}_{s,1}\left(r'_{i},\vec{\Lambda}'_{s,1}\right)\right\Vert \right.\nonumber \\
 & \left.\left.\times\int_{0}^{\infty}k'_{i,f}dk'_{i,f}\frac{2\hbar}{\pi}\cos\theta_{inc}\right)\right]\hat{R'}_{s,1}\cdot d\vec{S}_{sphere},\label{eq:sphere-ambient-force-average-TOTAL}\end{align}
 where $d\vec{S}_{sphere}$ is a differential surface element of a
sphere and the integration $\int_{S}$ is over the spherical surface.
The term $\vec{R'}_{s,0}$ is the initial crossing point inside the
sphere as defined in equation (\ref{eq:initial-arbitrary-k-vector-position-NA}).
The notation $\sum_{\left\{ \vec{k'}_{i,b},\vec{k'}_{i,f},\vec{R'}_{s,0}\right\} }$
imply the summation over all initial wave vector directions for both
inside $\left(\vec{k'}_{i,b}\right)$ and outside $\left(\vec{k'}_{i,f}\right)$
of the sphere, over all crossing points given by $\vec{R'}_{s,0}.$ 

It is easy to see that $\vec{\mathcal{F}}_{s,avg}$ of equation (\ref{eq:sphere-ambient-force-average})
is an {}``unregularized'' \textbf{1D} Casimir force expression for
the parallel plates (see the vacuum pressure approach by Milonni,
Cook and Goggin \cite{key-Milonni-Cook-Goggin}). It becomes more
apparent with the substitution $\triangle t=d/c.$ An application
of the Euler-Maclaurin summation formula \cite{key-Euler-Sum-Formula,key-Euler-Sum-Formula-Derivation}
leads to the regularized, finite force expression. The force $\vec{\mathcal{F}}_{s,avg}$
is attractive because \begin{align*}
\cos\theta_{inc} & >0\end{align*}
 and \begin{align*}
 & \sum_{n=1}^{\infty}\frac{n\pi}{\left\Vert \vec{R}_{s,2}\left(r'_{i},\vec{\Lambda}'_{s,2}\right)-\vec{R}_{s,1}\left(r'_{i},\vec{\Lambda}'_{s,1}\right)\right\Vert }\\
 & <\frac{1}{\pi}\lim_{r'_{i}\rightarrow\infty}\left(\left\Vert \vec{R}_{s,2}\left(r'_{i},\vec{\Lambda}'_{s,2}\right)-\vec{R}_{s,1}\left(r'_{i},\vec{\Lambda}'_{s,1}\right)\right\Vert \right.\\
 & \left.\times\int_{0}^{\infty}k'_{i,f}dk'_{i,f}\right),\end{align*}
 where $\left\Vert \vec{R}_{s,2}\left(r'_{i},\vec{\Lambda}'_{s,2}\right)-\vec{R}_{s,1}\left(r'_{i},\vec{\Lambda}'_{s,1}\right)\right\Vert $
is a constant for a given initial wave $\vec{k'}_{i,b}$ and the initial
crossing point $\vec{R'}_{s,0}$ in the cross-section of a sphere
(or hemisphere). The total average force $\vec{F}_{s,total},$ which
is really the sum of $\vec{\mathcal{F}}_{s,avg}$ over all $\vec{R'}_{s,0}$
and all initial wave directions, is therefore also attractive. For
the sphere configuration of Figure \ref{cap:sphere-reflection-dynamics},
where the energy flow direction is not restricted to the direction
of local surface normal, the Casimir force problem becomes an extension
of infinite set of parallel plates of a unit area.

\subsubsection{Hemisphere-Hemisphere and Plate-Hemisphere}

For the hemisphere-hemisphere and plate-hemisphere configurations,
the expression for the total average force per unit time, per initial
wave vector direction, per unit area is identical to that of the hollow
spherical shell with modifications, \begin{align}
\vec{\mathcal{F}}_{h,avg} & =\hat{R'}_{h,1}\left[\sum_{n=1}^{\infty}\frac{n\pi}{\left\Vert \vec{R}_{h,2}\left(r'_{i},\vec{\Lambda}'_{h,2}\right)-\vec{R}_{h,1}\left(r'_{i},\vec{\Lambda}'_{h,1}\right)\right\Vert }\right.\nonumber \\
 & -\frac{1}{\pi}\lim_{r'_{i}\rightarrow\infty}\left(\left\Vert \vec{R}_{h,2}\left(r'_{i},\vec{\Lambda}'_{h,2}\right)-\vec{R}_{h,1}\left(r'_{i},\vec{\Lambda}'_{h,1}\right)\right\Vert \right.\nonumber \\
 & \left.\left.\times\int_{0}^{\infty}k'_{i,f}dk'_{i,f}\right)\right]2\hbar\cos\theta_{inc},\label{eq:hemisphere-hemisphere-ambient-force-average}\end{align}
 where $\theta_{inc}\leq\pi/2$ and $n=1,2,3,\cdots.$ The incidence
angle $\theta_{inc}$ is from equation (\ref{eq:angle-of-incidence-exp});
$\vec{R}_{h,1}\left(r'_{i},\vec{\Lambda}'_{h,1}\right)$ and $\vec{R}_{h,2}\left(r'_{i},\vec{\Lambda}'_{h,2}\right)$
follow the generic form shown in equation (\ref{eq:Points-on-Hemisphere-R-NA}).
This force is attractive for the same reasons as discussed previously
for the hollow spherical shell case. The total radiation force averaged
over unit time, over all possible initial wave vector directions,
acting on the uncharged conducting hemisphere-hemisphere (plate-hemisphere)
surface is given by \begin{align}
\vec{F}_{h,total} & =\sum_{\left\{ \vec{k'}_{i,b},\vec{k'}_{i,f},\vec{R'}_{h,0}\right\} }\int_{S}\nonumber \\
 & \times\left[\sum_{n=1}^{\infty}\frac{2n\pi\hbar\cos\theta_{inc}}{\left\Vert \vec{R}_{h,2}\left(r'_{i},\vec{\Lambda}'_{h,2}\right)-\vec{R}_{h,1}\left(r'_{i},\vec{\Lambda}'_{h,1}\right)\right\Vert }\right.\nonumber \\
 & -\lim_{r'_{i}\rightarrow\infty}\left(\left\Vert \vec{R}_{h,2}\left(r'_{i},\vec{\Lambda}'_{h,2}\right)-\vec{R}_{h,1}\left(r'_{i},\vec{\Lambda}'_{h,1}\right)\right\Vert \right.\nonumber \\
 & \left.\left.\times\int_{0}^{\infty}k'_{i,f}dk'_{i,f}\frac{2\hbar}{\pi}\cos\theta_{inc}\right)\right]\hat{R'}_{h,1}\cdot d\vec{S}_{hemisphere},\label{eq:hemisphere-hemisphere-ambient-force-average-TOTAL}\end{align}
 where $d\vec{S}_{hemisphere}$ is now a differential surface element
of a hemisphere and the integration $\int_{S}$ is over the surface
of the hemisphere. The term $\vec{R'}_{h,0}$ is the initial crossing
point of the hemisphere opening as defined in equation (\ref{eq:initial-arbitrary-k-vector-position-NA}).
The notation $\sum_{\left\{ \vec{k'}_{i,b},\vec{k'}_{i,f},\vec{R'}_{h,0}\right\} }$
imply the summation over all initial wave vector directions for both
inside $\left(\vec{k'}_{i,b}\right)$ and outside $\left(\vec{k'}_{i,f}\right)$
of the hemisphere-hemisphere (or the plate-hemisphere) resonator,
over all crossing points given by $\vec{R'}_{h,0}.$ 

It should be remarked that for the plate-hemisphere configuration,
the total average radiation force remains identical to that of the
hemisphere-hemisphere configuration only for the case where the gap
distance between plate and the center of hemisphere is more than the
hemisphere radius $r'_{i}.$ When the plate is placed closer, the
boundary quantization length $\left\Vert \vec{L}\right\Vert $ must
be chosen carefully to be either \begin{align*}
\left\Vert \vec{L}\right\Vert  & =\left\Vert \vec{R}_{h,2}\left(r'_{i},\vec{\Lambda}'_{h,2}\right)-\vec{R}_{h,1}\left(r'_{i},\vec{\Lambda}'_{h,1}\right)\right\Vert \end{align*}
 or \begin{align*}
\left\Vert \vec{L}\right\Vert  & =\left\Vert \vec{R}_{p}\left(r'_{i},\vec{\Lambda}'_{p}\right)-\vec{R}_{h,N_{h,max}}\left(r'_{i},\vec{\Lambda}'_{h,N_{h,max}}\right)\right\Vert .\end{align*}
 They are illustrated in Figure \ref{cap:plate-hemisphere-plane-of-incidence-intersect-Complex}.
The proper one to use is the smaller of the two. Here $\vec{R}_{p}\left(r'_{i},\vec{\Lambda}'_{p}\right)$
is from equation (\ref{eq:Points-on-Plate-R-Final-NA}) of Appendix
C3 and $N_{h,max}$ is defined in equation (\ref{eq:N-max-Hemisphere-NA})
of Appendix C2.

\subsection{Interpretation of the Result}

Because only the specification of boundary is needed in Casimir's
vacuum-field approach as opposed to the use of a polarizability parameter
in Casimir-Polder interaction picture, the Casimir force is sometimes
regarded as a configurational force. On the other hand, the Casimir
effect can be thought of as a macroscopic manifestation of the retarded
van der Waals interaction. And the Casimir force can be equivalently
approximated by a summation of the constituent molecular forces employing
Casimir-Polder interaction. This practice inherently relies on the
material properties of the involved conductors through the use of
polarizability parameters. In this respect, the Casimir force can
be regarded as a material dependent force. 

Boyer's material property is such that the atoms in his conducting
sphere are arranged in such manner to respond only to the impinging
radiation at local normal angle of incidence to the sphere surface,
and they also radiate only along the direction of local normal to
its surface. When the Casimir force is calculated for a sphere made
of Boyer's fictitious material, the force is repulsive. Also, in Boyer's
original work, the laboratory boundary did not exist. When Boyer's
sphere is placed in a realistic laboratory, the net Casimir force
acting on his sphere becomes attractive because the majority of the
radiation from the laboratory boundaries acts to apply inward pressure
on the external surface of sphere when the angle of incidence is oblique
with respect to the local normal. If the sphere is made of ordinary
perfect conductors, the impinging radiation at oblique angles of incidence
would be reflected. In such cases the total radiation pressure applied
to the external local-sphere-surface is twice the pressure exerted
by the incident wave, which is the force found in equation (\ref{eq:sphere-ambient-force-average-TOTAL})
of the previous section. However, Boyer's sphere cannot radiate along
the direction that is not normal to the local-sphere-surface. Therefore,
the total pressure applied to Boyer's sphere is half of the force
given in equation (\ref{eq:sphere-ambient-force-average-TOTAL}) of
the previous section. This peculiar incapability of emission of a
Boyer's sphere would lead to the absorption of the energy and would
cause a rise in the temperature for the sphere. Nonetheless, the extra
pressure due to the waves of oblique angle of incidence is large enough
to change the Casimir force for Boyer's sphere from being repulsive
to attractive. The presence of the laboratory boundaries only act
to enhance the attractive aspect of the Casimir force on a sphere.
The fact that Boyer's sphere cannot irradiate along the direction
that is not normal to the local-sphere-surface, whereas ordinary perfect
conductors irradiate in all directions, implies that his sphere is
made of extraordinarily hypothetical material, and this may be the
reason why the repulsive Casimir force have not been experimentally
observed to date. 

In conclusion, (1) the Casimir force is both boundary and material
property dependent. The particular shape of the conductor, e.g. sphere,
only introduces the preferred direction for radiation. For example,
radiations in direction normal to the local surface has bigger magnitude
than those radiating in other directions. This preference for the
direction of radiation is intrinsically connected to the preferred
directions for the lattice vibrations. And, the characteristic of
lattice vibrations is intrinsically connected to the property of material.
(2) Boyer's sphere is made of extraordinary conducting material, which
is why his Casimir force is repulsive. (3) When the radiation pressures
of all angles of incidence are included in the Casimir force calculation,
the force is attractive for charge-neutral sphere made of ordinary
perfect conductor. And, lastly, (4) the Casimir force problem involving
any non-planar geometric boundary configurations can always be reduced
to a series of which the individual terms in the series are that of
the parallel plates problem. Because each terms in the series are
that of parallel plates problem, the summation over these individual
terms result in an attractive Casimir force.

\subsection{Suggestions on the Detection of Repulsive Casimir Force for a Sphere}

\begin{figure}
\begin{center}\includegraphics[%
  scale=0.5]{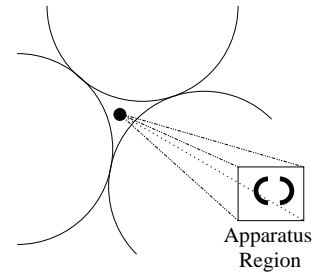}\end{center}

\caption{To deflect away as much possible the vacuum-field radiation emanating
from the laboratory boundaries, the walls, floor and ceiling are constructed
with some optimal curvature to be determined. The apparatus is then
placed within the {}``Apparatus Region.'' \label{cap:casimir-effect-test-lab}}
\end{figure}

The first step in detecting the repulsive Casimir force for a spherical
configuration is to find a conducting material that most closely resembles
the Boyer's material to construct two hemispheres. It has been discussed
previously that even Boyer's sphere can produce attractive Casimir
force when the radiation pressures due to oblique incidence waves
are included in the calculation. Therefore, the geometry of the laboratory
boundaries have to be chosen to deflect away as much as possible the
oblique incident wave as illustrated in Figure \ref{cap:casimir-effect-test-lab}.
Once these conditions are met, the experiment can be conducted in
the region labeled {}``Apparatus Region'' to observe Boyer's repulsive
force. 

\subsection{Outlook}

The Casimir effect has influence in broad range of physics. Here,
we list one such phenomenon known as {}``sonoluminescense,'' and,
finally conclude with the Casimir oscillator.

\subsubsection{Sonoluminescense}

The phenomenon of sonoluminescense remains a poorly understood subject
to date \cite{key-Gaitan-Crum-Church-Roy-Sono-Theory,key-Eberlein}.
When a small air bubble of radius $\sim10^{-3}\, cm$ is injected
into water and subjected to a strong acoustic field of $\sim20\, kHz$
under pressure roughly $\sim1\, atm,$ the bubble emits an intense
flash of light in the optical range, with total energy of roughly
$\sim10^{7}\, eV.$ This emission of light occurs at minimum bubble
radius of roughly $\sim10^{-4}\, cm.$ The flash duration has been
determined to be on the order of $100\, ps$ \cite{key-Gompf-Gunther-Nick-Pecha-Eisen-Sono-Measure,key-Hiller-Putterman-Weninger-Sono-measurement,key-Moran-Sweider-Sono-measure}.
It is to be emphasized that small amounts of noble gases are necessary
in the bubble for sonoluminescense. 

\begin{figure}
\begin{center}\includegraphics[%
  scale=0.5]{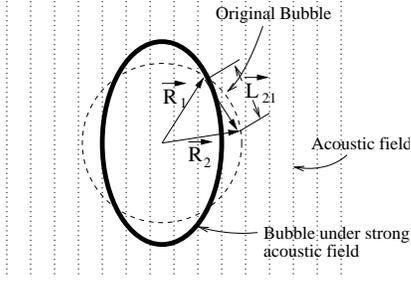}\end{center}

\caption{The original bubble shape shown in dotted lines and the deformed
bubble in solid line under strong acoustic field. \label{cap:sonoluminescense}}
\end{figure}

The bubble in sonoluminescense experiment can be thought of as a deformed
sphere under strong acoustic pressure. The dynamical Casimir effect
arises due to the deformation of the shape; therefore, introducing
a modification to $\vec{L}_{21}=\left\Vert \vec{R}_{2}-\vec{R}_{1}\right\Vert $
from that of the original bubble shape. Here $\vec{L}_{21}$ is the
path length for the reflecting wave in the original bubble shape.
In general $\vec{L}_{21}\equiv\vec{L}_{21}\left(t\right)=\left\Vert \vec{R}_{2}\left(r_{i}\left(t\right),\theta\left(t\right),\phi\left(t\right)\right)-\vec{R}_{1}\left(r_{i}\left(t\right),\theta\left(t\right),\phi\left(t\right)\right)\right\Vert .$
From the relations found in this work for the reflection points $\vec{R}_{1}\left(r_{i}\left(t\right),\theta\left(t\right),\phi\left(t\right)\right)$
and $\vec{R}_{N}\left(r_{i}\left(t\right),\theta\left(t\right),\phi\left(t\right)\right),$
together with the dynamical Casimir force expression of equation (\ref{eq:dynamical-force-L-dot-ONLY-3D-NA}),
the amount of initial radiation energy converted into heat energy
during the deformation process can be found. The bubble deformation
process shown in Figure \ref{cap:sonoluminescense} is a three dimensional
heat generation problem. Current investigation seeks to determine
if the temperature can be raised sufficiently to cause deuterium-tritium
(d-t) fusion to occur, which could provide an alternative approach
to achieve energy generation by this d-t reaction (threshold $\sim17\, KeV$)
\cite{key-A-Prosperetti}. Its theoretical treatment is similar to
that discussed on the \textbf{1D} problem shown in Figure \ref{cap:casimir-plates-outlook}. 

\subsubsection{Casimir Oscillator}

If one can create a laboratory as shown in Figure \ref{cap:casimir-effect-test-lab},
and place in the laboratory hemispheres made of Boyer's material,
then the hemisphere-hemisphere system will execute an oscillatory
motion. When two such hemispheres are separated, the allowed wave
modes in the hemisphere-hemisphere confinement would no longer follow
Boyer's spherical Bessel function restriction. Instead it will be
strictly constrained by the functional relation of $\left\Vert \vec{R}_{2}-\vec{R}_{1}\right\Vert ,$
where $\vec{R}_{1}$ and $\vec{R}_{2}$ are two neighboring reflection
points. Only when the two hemispheres are closed, would the allowed
wave modes obey Boyer's spherical Bessel function restriction. 

\begin{figure}
\begin{center}\includegraphics[%
  scale=0.4]{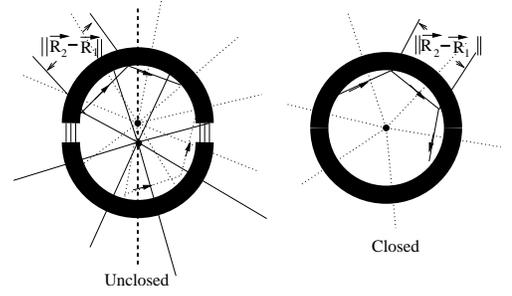}\end{center}

\caption{The vacuum-field radiation energy flows are shown for closed and
unclosed hemispheres. For the hemispheres made of Boyer's material,
the non-radial wave would be absorbed by the hemispheres. \label{cap:cho-casimir-sphere}}
\end{figure}

Assuming that hemispheres are made of Boyer's material and the laboratory
environment is that shown in Figure \ref{cap:casimir-effect-test-lab},
the two closed hemispheres would be repulsing because Boyer's Casimir
force is repulsive. Once the two hemispheres are separated, the allowed
wave modes are governed by the internal reflections at oblique angle
of incidence. Since the hemispheres made of Boyer's material are {}``infinitely
unresponsive'' to oblique incidence waves, all these temporary non-spherical
symmetric waves would be absorbed by the Boyer's hemispheres and the
hemispheres would heat up. The two hemispheres would then attract
each other and the oscillation cycle repeats. Such a mechanical system
may have application. 

\begin{acknowledgments}
I am grateful to Professor Luke W. Mo, from whom I have received so
much help and learned so much physics. The continuing support and
encouragement from Professor J. Ficenec and Mrs. C. Thomas are gracefully
acknowledged. Thanks are due to Professor T. Mizutani for fruitful
discussions which have affected certain aspects of this investigation.
Finally, I express my gratitude for the financial support of the Department
of Physics of Virginia Polytechnic Institute and State University.
\end{acknowledgments}

\end{document}